\documentclass[twocolumn,trackchanges]{aastex631}
\usepackage{amsmath}
\usepackage{graphicx}
\usepackage{textcomp}
\usepackage{soul}

\shorttitle{AASTeX v6.3.1 Sample article}
\shortauthors{Margoti et al.}
\graphicspath{{./}{figures/}}

\begin{document}

\title{Ellipsoidal Modeling Framework for TNOs: the triaxial shape of (50000) Quaoar and its implications for Weywot and the dynamical environment}

\author[0000-0002-2103-4408]{Giuliano Margoti}
\affiliation{Observatório Nacional/MCTI, Rio de Janeiro, Brazil}
\affiliation{Laboratório Interinstitucional de e-Astronomia – LIneA, Av. Pastor Martin Luther King Jr 126, Rio de Janeiro, RJ 20765-000, Brazil}

\author[0000-0003-2311-2438]{Felipe Braga-Ribas}
\affiliation{Federal University of Technology - Paraná (PPGFA/UTFPR-Curitiba), Av. Sete de Setembro, 3165, CEP 80230-901 - Curitiba - PR - Brazil}
\affiliation{Observatório Nacional/MCTI, Rio de Janeiro, Brazil}
\affiliation{Laboratório Interinstitucional de e-Astronomia – LIneA, Av. Pastor Martin Luther King Jr 126, Rio de Janeiro, RJ 20765-000, Brazil}

\author[0009-0001-5754-4688]{Eros Gradovski}
\affiliation{Observatório Nacional/MCTI, Rio de Janeiro, Brazil}
\affiliation{Laboratório Interinstitucional de e-Astronomia – LIneA, Av. Pastor Martin Luther King Jr 126, Rio de Janeiro, RJ 20765-000, Brazil}


\author[0000-0001-5138-230X]{Gustavo Madeira }
\affiliation{Observatório Nacional/MCTI, Rio de Janeiro, Brazil}

\author[0000-0003-1000-8113]{Chrystian L. Pereira}
\affiliation{Observatório Nacional/MCTI, Rio de Janeiro, Brazil}
\affiliation{Laboratório Interinstitucional de e-Astronomia – LIneA, Av. Pastor Martin Luther King Jr 126, Rio de Janeiro, RJ 20765-000, Brazil}

\author[0000-0002-8690-2413]{Jose L. Ortiz}
\affiliation{Instituto de Astrofísica de Andalucía (IAA-SIC), Glorieta de la Astronomía s/n, 18008 Granada, Spain}

\author[0000-0003-1995-0842]{Bruno Sicardy}
\affiliation{LESIA, Observatoire de Paris, Université PSL, CNRS, Sorbonne Université, 5 place Jules Janssen, 92190 Meudon, France}



\begin{abstract}

The three-dimensional shapes of Trans-Neptunian Objects (TNOs) are notoriously difficult to determine due to their slow orbital motion and minimal variations in aspect angle over human timescales. To overcome this geometric degeneracy, we developed a versatile modeling framework capable of deriving ellipsoidal shapes by simultaneously combining stellar occultations and rotational light curves, including parameterizations of localized surface albedo. We applied this methodology to the large TNO (50000) Quaoar, jointly modeling 37 stellar occultation events and two photometric datasets. Our MCMC analysis demonstrates that Quaoar is best represented by a triaxial shape with equatorial semi-axes $a = 620.7^{+3.3}_{-2.8}$ km and $b = 546.5^{+6.0}_{-5.6}$ km, and a polar semi-axis $c = 511.5^{+5.3}_{-5.5}$ km, yielding a bulk density of $1.663 \pm 0.127$ g cm$^{-3}$. 
By isolating Quaoar's barycentric motion from the astrometric offsets, we determined the mass of its satellite, Weywot, to be $(1.485 \pm 0.360) \times 10^{18}$ kg, corresponding to a system mass ratio of $q = 0.0012 \pm 0.0003$ and a satellite bulk density of $0.95 \pm 0.60$ g cm$^{-3}$.
Furthermore, dynamical simulations using the derived triaxial gravitational field reveal that potential additional satellites can only maintain stable orbits beyond 2.4 Quaoar radii, as inner bodies would be strongly perturbed. Regarding the ring system, particles at the location of the secondary ring (Q2R) location exhibit libration associated with the 5:7 spin-orbit resonance (SOR). Similarly, particles at the location of the main ring (Q1R) location show libration of the 2:6 SOR angle for orbital eccentricities larger than 0.2, providing a plausible mechanism for the azimuthal width variations observed in Q1R.

\end{abstract}



\section{Introduction} \label{Introduction}

Most of the large Trans-Neptunian Objects (TNOs) were discovered in the early 2000s, therefore, only a small fraction of their orbits have been observed so far. Consequently, the variation in the aspect angle of these bodies is minimal over a typical 20 to 30-year study period, meaning we observe essentially the same aspect angle throughout this interval \citep{Sicardy_2024}.

This characteristic limits not only the determination of the object's orbit but also its shape \citep{Margoti_2024}. Furthermore, the stellar environment can hinder these studies; for instance, if an object is located in an extremely dense stellar region, high-quality light curves or spectra may be unattainable due to flux contamination from nearby stars. This is the case for objects such as Ixion \citep{Kilic_2026}, Mani \citep{Rommel_2023}, 2012 KX14 \citep{Rizos_2025}, and currently is also the case for Quaoar.

In this work, we utilize rotational light curve observations and stellar occultations, collected from \citet{Ortiz_2003}, \citet{Proudfoot_2025}, and \citet{Margoti_2026}, to determine the three-dimensional shape of (50000) Quaoar, an object that exhibits a small aspect angle variation of 1.7°, as well as a small variation in its pole position angle of 17.8°, across the observational epochs analyzed in this study. 

Light curves involve direct imaging of the object over long periods to record its brightness variations. For small, irregular objects, surface topography dominates the light curve morphology (e.g. \citet{Gradovski_2026}). However, for TNOs with absolute magnitudes between 4 and 8 (corresponding to radii larger than 100 km), they are expected to be in hydrostatic equilibrium (e.g. \citet{Grundy_2019}). In this case, topographic variations are not significant enough to cause major changes in the light curve \citep{Margoti_2024}. An object in hydrostatic equilibrium will assume an ellipsoidal shape defined by three semi-axes: the primary equatorial semi-axis (a), the secondary equatorial semi-axis (b), and the polar semi-axis (c).

The light curve will exhibit distinct characteristics, provided it is not contaminated by a satellite. If the object is spherical ($a=b=c$) or oblate ($a=b>c$), the projected area on the sky remains constant, resulting in a flat light curve. However, as observed on Pluto and Charon by the New Horizons mission, large objects may possess significant surface heterogeneity resulting in albedo variations (e.g. \citet{Stern_2015, Grundy_2016_2}). Such variations, when observed during rotation, introduce brightness changes, turning a flat curve into a single-peaked light curve (one peak and one valley per rotation period).

For a triaxial object ($a>b>c$), the brightness varies due to changes in the projected area, presenting the observer with the largest area twice per full rotation, which results in a double-peaked light curve. If the shape is the sole driver, both peaks will have equal heights. If an albedo variation is also present, it will cause an additional brightness change, making one of the peaks or valleys higher or lower, as the spot-induced variation occurs only once per rotation, as seen in the cases of Haumea \citep[e.g.,][]{Lacerda_2008} and (84522) 2003~VS$_2$ \citep[e.g.,][]{Rossi_2019, Lubiano_2022}. While the amplitude of a ``pure'' triaxial shape defines the axes ratio and possible pole orientations, albedo contamination prevents such a direct determination.

Stellar occultations are the passage of a Solar System object in front of a background star for a specific observer, and provide a "snapshot" of the object’s instantaneous limb \citep[e.g.,][]{Sicardy_2024}. Multiple observers detecting the same event enable the precise capture of the shape at that moment. Successive occultations act as a form of ``tomography'', enabling the reconstruction of a mean ellipsoid \citep[e.g.,][]{Margoti_2024, Gradovski_2026}.

The difficulty lies in the connection between these methods. In dense stellar fields, occultation candidates are abundant, but rotation light curves are hindered by contamination, as seen in the cases of Quaoar \citep{Margoti_2026}, Chariklo \citep{Morgado_2021}, and Mani \citep{Rommel_2023}. Conversely, in sparse fields, light curves are easier to obtain, but occultation candidates are scarce, as is the case of Haumea \citep{Ortiz_2017}. While occultations provide high precision for the projected limb, variables such as pole orientation and rotation period remain sub-determined. A single light curve, however, can accurately determine the rotation period but impose only weak constraints on the semi-axes and pole projection. By combining both methods, we achieve a more complete determination of the ellipsoidal shape and its rotational characteristics \citep[e.g.,][]{Gradovski_2026}. Furthermore, centers derived from joint analysis are more reliable for orbital improvement \citep{Desmars_2015}.


\textcolor{black}{The strategy of combining rotational light curves and stellar occultations to reconstruct 3D shape models is a well-established and highly successful technique, extensively applied to Main Belt asteroids over the past decades \citep[e.g.,][]{Bartczak_2018, Viikinkoski_2015, Durech_2011, Durech_2010}. However, when applied to the outer Solar System, methodologies often have to rely on specific geometric or observational constraints.} For instance, \citet{Gradovski_2026} derived the tridimensional shape for irregular Trojans using occultation and rotational data, but their uncertainty determination was based on evaluating the shape model obtained previously. \citet{Morgado_2021} fitted a triaxial model for Chariklo, and \citet{Leiva_2017} explored spherical, oblate, and triaxial solutions, however, neither study incorporated rotational light curves to constrain the 3D shape. More recently, \citet{Proudfoot_2026}, based on \citet{margoti2024}, presented a 3D modeling approach that utilizes rotational data, but it is limited to single-epoch occultation events and relies on fixing both the rotational amplitude and the object's pole orientation.

In contrast, our framework operates without the need for restrictive priors. It simultaneously integrates multi-epoch stellar occultations and rotational light curves, with the flexibility to also incorporate phase curve data. By mapping the exact spatial coordinates of each structural facet, our model can naturally parameterize localized albedo variegations or prominent topographic features. Most importantly, unlike the approach by \citet{Gradovski_2026}, we do not evaluate the uncertainties of a fixed geometry; instead, our algorithm dynamically determines the 3D shape model and its intrinsic uncertainties concurrently.

Following this introduction, Section \ref{sec: Metodology} details the Methodology: Shape representation (Section \ref{sec:2.1methods}), orientation (Section \ref{sec: Orientation}), rotational light curve modeling (Section \ref{sec: Rotational light curve}), stellar occultation fitting (Section \ref{sec: Stellar occultation}), and the overall modeling process (Section \ref{sec: Modeling process}). In Section \ref{sec: Application}, we apply this methodology to Quaoar, calculating a new rotation period and deriving a pure triaxial shape. In Section \ref{sec: Resultados}, we present the results of this analysis, encompassing the shape modeling (Section \ref{sec: Modeling}), our interpretation of the results to obtain Weywot's mass (Section \ref{weywotmass}), and the dynamical implications for the Quaoar environment (Section \ref{sec:environment}). Finally, we provide our concluding remarks in Section \ref{sec:conclusions}. 

Additionally, we include two appendices in this work. In Appendix \ref{ap:albedo}, we present the methodology for incorporating an albedo variation into our models, applying it to the model published by \citet{Margoti_2026}, and in Appendix \ref{sec:New astrometric position}, we provide the new astrometric positions derived from the model presented in the main part of this work.

\section{Methodology}\label{sec: Metodology}

Typically, inner asteroid models are constructed by a mesh of vertices in 3D space connected by triangular faces. This approach is advantageous because it allows the construction of various convex or non-convex shapes. Moreover, these objects are small and observed from multiple aspect angles in just a few years. This produces light curve amplitude and form variation due to their surface irregularities, enabling the model solution to be quite irregular  \citep[e.g.,][]{Bartczak_2018}.

In the case of TNOs, the situation is somewhat different. Typically, only two or three light curves are available, all recorded from similar aspect angles, even if captured over long period intervals \citep{Sicardy_2024}. This is because they are among the most distant objects observed in the Solar System, with orbital periods exceeding 200 years. As a result, light curves alone are insufficient to reconstruct the irregular shapes of these bodies \citep[e.g.,][]{Ortiz_2003}. Furthermore, most of the known TNOs are larger than 200 kilometers, which contributes to a more regular shape \citep[e.g.,][]{Gradovski_2026}.

\subsection{Shape representation}\label{sec:2.1methods}

Our analysis will be limited to determining the size of the three axes of the body, the primary equatorial axis ($a$), the secondary equatorial axis ($b$) and the polar axis ($c$), yielding only three possible shapes: spherical ($a = b = c$), oblate ($a = b > c$) or triaxial ($a > b > c$).  

To construct our surface model, we implemented the following procedure: we created a parametric grid with 100 longitude lines and 100 colatitude lines. Using these coordinates and the lengths of the three semi-axes, we generated the vertices of an ellipsoid in Cartesian space. These vertices were then connected to form a mesh composed of triangular faces. To allow for the subsequent application of the rotation curve model, we calculated the geometric center of each triangular face and converted these Cartesian coordinates back into spherical latitude and longitude values. This procedure provides a fully mapped structural model that can be tested against our combined stellar occultation and rotation data.

\subsection{Orientation}\label{sec: Orientation}

To define both the rotation curve and the object's limb at a specific time, it is necessary to determine the sub-observer point: the location on the object's surface (specified by latitude and longitude) where the observer would appear at zenith. For all purposes, we adopt the nomenclature derived from \cite{Archinal_2018}.

Determining the sub-observer point requires, above all, knowledge of the object's pole orientation, specifically, its spin axis (fixed as the polar axis), as well as its rotation period and initial phase. In our context, zero latitude and zero longitude indicate that the secondary equatorial semi-axis is oriented toward the observer.

The model is defined within its own reference frame, where the z-axis lies along the primary equatorial axis and the x-axis along the secondary equatorial axis. To orient the sub-observer point, it is necessary to know the Euler angles $\zeta$, $\varphi$ and $\xi$ which are applied in a 3–1–3 rotation sequence about the z-, x-, and z-axes, respectively.

To combine the translation (orbital position) with the object's rotation, we use a 3×3 matrix representing the sub-observer point:
\begin{equation}\label{eq: matriz M}
    M = T R_z(\zeta) R_x(\varphi) R_z(\xi) ,
\end{equation}
where $\zeta = \alpha_0 + 90º$, $\varphi = 90 - \delta_0$ and $\xi = W$, with $\alpha_0$ is the right ascension of the pole, $\delta_0$ the declination of the pole, and $W$ the 
\textcolor{black}{current rotational phase} of the object. In this formulation, $M$ represents the sub-observer point in distance units $(x_{sub}, y_{sub}, z_{sub})$, while $T$ is the translation vector giving the object's center coordinates $(x, y, z)$ e.g relative to the geocenter, in distance units.

In this work, we assume that the rotational spin is aligned with the polar axis. The primary equatorial axis is chosen as the reference point on the object's surface to define the phase. Therefore, to determine the object's phase at any given instant, we use the equation
\begin{equation}\label{eq: W}
    W = W_0 + \dot{W}(t-t_{ref}),
\end{equation}
where $W$ is the phase in this instant $t$, $W_0$ is the initial phase in a reference instant $t_{ref}$ and $\dot{W}$ is the angular speed .

To obtain the latitude and longitude of the sub-observer point, the following procedure is used:
\begin{equation}\label{eq: lon, lat, r}
    \begin{split}
        r = \sqrt{x_{sub}^2+y_{sub}^2+z_{sub}^2}; \\
        lon = \arctan{\frac{y_{sub}}{x_{sub}}}; \\
        lat = \arccos{z_{sub}/r}. \\
    \end{split}
\end{equation}

Now, to determine how the object is oriented in the sky plane, it is necessary to use the sub-observer point and the projection’s position angle:
\begin{equation}\label{eq: PA}
    PA = \arctan{\frac{\sin\Delta\alpha}{\cos{\delta}\tan{\delta_0} - \sin{\delta}\cos{\Delta\alpha}}}
\end{equation}
where $\alpha$ and $\delta$  are, respectively, the right ascension and declination from the object's ephemeris, and $\Delta\alpha$ is the difference between $\alpha_0$ and $\alpha$

Finally, the projection of each point of the shape onto the sky plane is determined by equation \ref{eq: matriz M}, where $\zeta = lat$, $\varphi = PA$ and $\xi = lon$. In this way, the matrix $M$ will have its values in distance units, where values corresponding to $z > 0$ represent the visible portion of the object and $z = 0$ the limb of the object.

\subsection{Rotational light curve}\label{sec: Rotational light curve}

Due to the limited variation in aspect angle, TNO light curves are traditionally used to constrain the rotational period and the ratios of the body's semi-axes. To derive the rotation period of the object, we employ the Lomb-Scargle periodogram, which is particularly well-suited for analyzing unevenly spaced photometric data \citep{Lomb_1976, Scargle_1982}.

To model the rotational light curve, we compute the total reflected flux from the shape model at each observation epoch. The surface of our triaxial mesh is treated as a collection of triangular facets. The light reflected by each facet is simulated using a linear combination of the Lambert and Lommel-Seeliger scattering laws \citep[e.g.,][]{Kaasalainen2001}, given by:
\begin{equation}\label{eq: LLS}
    S = (1-c)\frac{\mu \mu_0}{\mu + \mu_0} + c\mu \mu_0,
\end{equation}
where $\mu$ and $\mu_0$ denote the cosines of the angles between the surface normal and the direction to the observer, and between the surface normal and the direction to the Sun, respectively. The linear weight factor $c$ is set to 0.1, representing a 10\% Lambertian light curve. A given facet only contributes to the total observable flux if it is both visible to the observer ($\mu > 0$) and illuminated by the Sun ($\mu_0 > 0$).

Finally, the integrated flux of the entire observable and illuminated surface is converted into a synthetic relative magnitude. 
\textcolor{black}{The object's ephemeris, including the projection's position angle and light-travel time corrections at the precise time of each observation, are obtained using the SORA library \citep{sora_2022}. These geometric parameters are then used as inputs for our custom 3D rotation and projection scheme described in Section 2.2.}
This synthetic curve is then compared against the observed photometric datasets. Since our observational data may originate from different campaigns with varying photometric precision, the goodness-of-fit is evaluated using a weighted sum of squared residuals ($\chi^2$), where the contribution of each dataset is weighted by its respective observational uncertainty ($\sigma$). This approach allows us to constrain the semi-axes and pole orientation for a homogeneous body.

\subsection{Stellar occultation}\label{sec: Stellar occultation}

A stellar occultation provides a set of points on an object's surface, the immersion and emersion points serve as a high-resolution 2D snapshot of the object's limb at the time of the event \citep[e.g.,][]{Sicardy_2024}. By selecting a reference time $t_{ref}$ and an initial rotational phase $W_0$, we can determine the specific rotational state of the body during the occultation.

To compare our 3D ellipsoid model with the occultation chords, we first compute the sub-observer coordinates and the pole position angle at the epoch of each observation, accounting for the light travel time between the object and the observer. Using the procedures described in Section \ref{sec: Orientation}, the 3D surface defined by the semi-axes $a$, $b$, and $c$ is rotated and projected onto the sky plane. The apparent limb of the model is then computationally extracted by dividing the projected shape into narrow angular bins and identifying the outermost projected points from the body's center.

An extra step in modeling stellar occultations is adjusting the astrometric center of the event in the sky plane ($f, g$), as uncertainties in the object's ephemeris can introduce systematic offsets. To address this, we implemented an optimization routine that applies random positional shifts, drawn from a normal distribution, to the model's center. For each iteration, we calculate the radial distances between the observed occultation extremities and the nearest points on the modeled limb, in other words, we marginalize the center position.

The optimal event center is defined as the one that minimizes these radial residuals. Finally, the goodness-of-fit for the occultation data is evaluated using a $\chi^2$ statistic. 

To account for small-scale topographic irregularities not captured by the triaxial ellipsoid, an intrinsic model uncertainty ($\sigma_{model}$ km) is added in quadrature to the individual observational uncertainties of the occultation chords. This value of $\sigma_{model}$ was estimated for a local radial topography based on the methodology proposed by \citet{Johnson_1973}:
\begin{equation} \label{eq: topography} 
  h \leq \frac{3kS}{4\pi G\rho^2 R}, 
\end{equation}
where $h$ represents the maximum topographic height that the object can sustain, $k$ is a constant that indicates the material's structural properties ($k=1$ for a homogeneous body, or $k>1$ for a differentiated structure with increased resistance toward the center), $S$ is the material's strength, $G$ is the gravitational constant, $\rho$ is the density of the object, and $R$ is its volumetric equivalent radius. 

\subsection{Modeling process}\label{sec: Modeling process}

To compare the model with the occultation and rotational points we use the $\chi^2$ method:
\begin{equation}\label{eq: chi2}
  \chi^2 = \sum_{i=1}^N \frac{(r_i - \bar{r}_i)^2}{\sigma_i^2 + \sigma_{model}^2}, 
\end{equation}
where $r_i$ is the radial distance of the $i$th observed chord or point in the rotational light curve, $\bar{r}_i$ is the corresponding radial distance on the ellipsoidal limb for occultation or the point in the sintetic light curve, $\sigma_i$ is the uncertainty in the observed radial distance or the uncertainty of the rotation point, and the term $\sigma_{model}$ accounts for a minimum uncertainty in the chords, ensuring that the final result is not dominated by local topographic features present in highly precise chords \citep{sora_2022}, we do not use $\sigma_{model}$ in the rotational light curve.

The modeling process consists of a series of logical steps to be followed. First, the rotation light curves of the object to be analyzed are obtained either from a repository or directly from the authors of the articles. It is important to verify whether the light curves were acquired without filters and whether they need to be corrected for light-travel time \citep{Gradovski_2026}. Once these aspects are confirmed, the Lomb-Scargle method is applied to determine the best rotation period. The period search should be performed considering both a single peak and a double peak; if the object is an oblate ellipsoid or spherical with an albedo variation on its surface, it will display only one peak throughout its rotation, whereas a triaxial shape will exhibit a double peak, possibly with albedo variations causing a difference in height between the peaks. This period will be used as the model's prior \citep{Margoti_2024}.

Then, based on the time instants and coordinates provided by the stellar occultation, the chords for each event are generated. Within the previously imposed limits, a search is then conducted for the best ellipsoid that represents the data, using both the rotational light curve and occultation, or only the occultation \citep{Margoti_2026}. For this, a genetic algorithm is used, in which two initial populations of 100 random ellipsoids each are generated and crossed to produce 10 new descendants, of which 2 from each are incorporated into two new groups of 100 ellipsoids that are crossed again. The crossover method consists of randomly shuffling the variables and allowing small mutations, thereby ensuring better exploration of the parameter space. At the end of 50 generations, the crossover is halted to select the best ellipsoid, defined as the one among all generations that achieved the lowest value of $\chi^2$ \citep{Margoti_2024}. This best ellipsoid is then used as the initial parameter for modeling with Markov Chain Monte Carlo (MCMC) Ensemble Sampler \citep{Mackey_2013} to ensure thorough exploration and proper determination of the posterior uncertainties of the modeled parameters. The constraints for parameter space exploration must be defined at this stage, based on the prior knowledge of the object. Specifically, we apply uniform priors that restrict the semi-axes ($a \ge b \ge c$), pole coordinates, rotation period, and initial phase within physically meaningful boundaries.

Because these physical parameters span different orders of magnitude, the parameter space is normalized to ensure efficient sampling. We deploy 50 independent walkers initialized around the best solution provided by the genetic algorithm. The MCMC is executed in two phases: an initial burn-in phase of 1000 steps to allow the Markov chains to thermalize, followed by a main sampling phase of 5000 steps. The marginal posterior distributions are then extracted from these final chains, with the best-fit values and their respective uncertainties defined by the 50th, 16th, and 84th percentiles.

It is important to note that the object's ephemeris has uncertainties, therefore the center of the projected ellipsoid's limb will not be aligned with the origin. Keeping that in mind, to avoid unnecessary tests, a search is performed for the optimal center of the limb, that is, the center that minimizes the value of $\chi^2$ \citep{Margoti_2026}.

To exhaust the modeling possibilities, one can choose to ignore the amplitude constraints imposed by the light curve and rely solely on the rotation period determined by the Lomb-Scargle method. In this scenario, the 3D shape is constrained exclusively by the occultation chords. This is a robust approach provided there is a sufficient temporal distribution of occultation events covering different rotational phases. Consequently, if the theoretical light curve amplitude predicted by this purely geometric model differs from the observed photometric amplitude, the discrepancy can be directly attributed to surface albedo variations. Furthermore, it is possible to impose specific geometric constraints on the model, such as assuming an oblate ($a=b$) or a spherical shape ($a=b=c$), as demonstrated by \citet{Margoti_2026}, \citet{Morgado_2021} and \citet{Leiva_2017}.

\section{Application}\label{sec: Application}

Four rotational light curves for Quaoar have been reported in the literature. The data set from \citep{Rabinowitz_2007} is available in the Vizier catalog, but we could not use it because the observations were obtained with different color filters, while the data from \citet{Thirouin_2010} and \citep{Kiss_2024} are also accessible via Vizier. Unfortunately, the light curves exhibit high dispersion, and they are unsuitable for our analysis. The data set from \citet{Ortiz_2003} provides the best rotational data for Quaoar. In this work, we reprocessed the images to improve the photometry with modern tools \citep{Assafin_2023} and derived new rotational light curves.

The data were filtered to ensure optimal quality, resulting in the selection of observations from May 23–24 and June 20–22, 2003. Detailed information on these data can be found in \citet{Ortiz_2003}. We performed the differential aperture photometry using PRAIA \citep{Assafin_2023} and converted the flux to relative magnitude using the equation:
\begin{equation}\label{eq: m_r}
     m_r = -2.5\log_{10}F,
\end{equation}
where $m_r$ is the relative magnitude, and in this case F is $\frac{F_i}{F_m}$ where $F_i$ is the flux ratio of Quaoar in the $i$th image, and $F_m$ is the Quaoar mean flux ratio of the entire dataset. To ensure consistency, the same field calibrators were used for all images.

\begin{figure*}[htb!]
\gridline{\fig{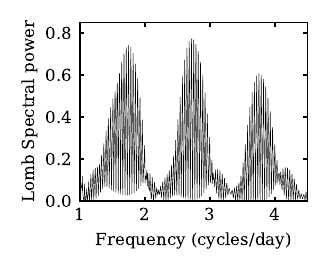}{0.4\textwidth}{(a)}
          \fig{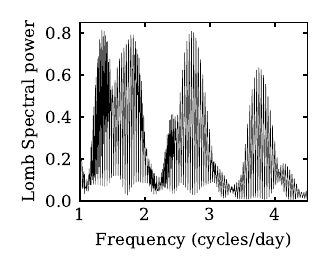}{0.4\textwidth}{(b)}
          }
\gridline{\fig{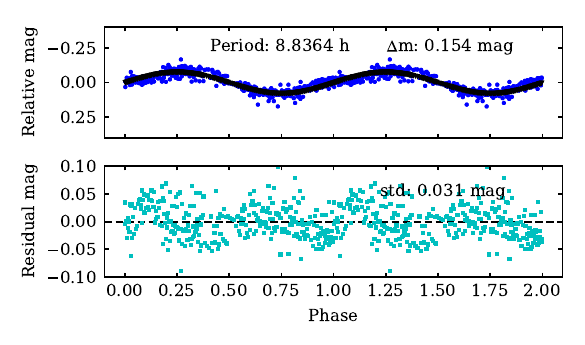}{0.5\textwidth}{(c)}
          \fig{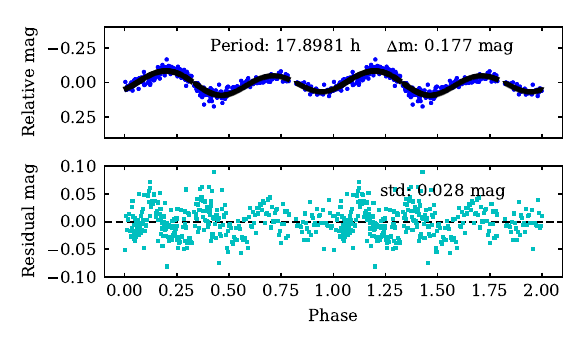}{0.5\textwidth}{(d)}
          }
\caption{Lomb-Scargle periodograms and folded light curves for single- and double-peaked. Panel (a) shows the periodogram of the combined data set for a single-peaked with 1‑sigma $8.8363 \pm 0.0017$ h and 3‑sigma $8.8364 \pm 0.0031$ h errors. Panel (b) similar to (a) but for a double peak the 1‑sigma uncertainty is ($17.8979 \pm 0.0031$ h) and 3‑sigma is ($17.8980 \pm 0.0060$ h). Panel (c) displays the light curve (blue dots) with the best frequency applied in the top section and the residuals from the Lomb–Scargle fit in the bottom section for a single-peak. Panel (f) similar to (a) but for a double peak.
}\label{fig: rot quaoar}
\end{figure*}

In the analyses we find a single-peak fit to the data resulted in a rotation period of $8.8364 \pm 0.0031$ hours, the periodogram of the combined sets is shown in Figure \ref{fig: rot quaoar} panels (a), and (c). However, as described by \citet{Ortiz_2003} and \citet{Kiss_2024}, Quaoar is likely triaxial. Its rotational light curve, assuming no albedo variation, would exhibit a double peak with a rotation period twice the initially derived value, at $17.6728 \pm 0.0062$ hours. These results are consistent with those presented by  \citet{Ortiz_2003}, \citet{Thirouin_2010}, \citet{Kiss_2024} and \citep{Rabinowitz_2007} as shown in table \ref{tab: rotational period by Quaoar}.

In addition to fitting a single peak, we can also consider peaks of different heights when fitting the period. This approach is not simply a matter of doubling the rotation period, which would be valid only for an object whose light curve is solely determined by changes in the reflecting area, but rather involves fitting a period that simultaneously accounts for both peaks, accounting for an irregular shape or albedo variegations. By doing so, we obtain the periodogram shown in Figure \ref{fig: rot quaoar} panels (b), and (d), which yields a rotation period of $17.8980 \pm 0.0060$~h, slightly larger than the period obtained by doubling the single peak period. In addition, the double peak has an amplitude of $0.177 \pm 0.060$, which is slightly higher than the single peak shown in Table \ref{tab: rotational period by Quaoar}; however, the error bars still overlap.

\begin{table*}[htb!]
\caption {Rotational period by Quaoar} \label{tab: rotational period by Quaoar}
\centering
\begin{tabular}{llll}
\hline\hline             
Single peak (h)& Double peak (h) & Amplitude (mag)& Reference \\
\hline
$8.8364 \pm 0.0031$ & $17.6728 \pm 0.0062$ & $0.154 \pm 0.030$ & This Work\\
                    & $17.8980 \pm 0.0060$ & $0.177 \pm 0.060$ & This Work\\
$8.8394 \pm 0.0002$ & $17.6788 \pm 0.0004$ & $0.133 \pm 0.028$ & \citet{Ortiz_2003}\\
$8.8400$ & $17.6800$ & $0.150 \pm 0.040$ & \citet{Thirouin_2010}\\
$8.8760 \pm 0.0720$ & $17.7520 \pm 0.1440$ & $0.154 \pm 0.040$ & \citet{Kiss_2024}\\
$8.8400 \pm 0.0083$ & $17.6800 \pm 0.0167$ & $0.180 \pm 0.100$ & \citet{Rabinowitz_2007}\\
\hline
\end{tabular}
\end{table*}

With these results, we establish our first prior for the model: the rotation period. Instead of arbitrarily imposing a value, we evaluated both the doubled single-peak period and the direct double-peak period. We adopted the doubled single-peak period for our primary results, as it yields the simplest pure triaxial model capable of consistently explaining both the rotational light curves and the stellar occultations. Models assuming the alternative double-peak period proved unfavorable when confronted with the HST rotation data discussed below, and are therefore not presented. Furthermore, the purely oblate scenario for Quaoar has already been extensively tested by \citet{Margoti_2026}. Finally, for the sake of completeness, we verify in Appendix \ref{ap:albedo} how an albedo variation could potentially drive the light curve behaviour shown in Figure \ref{fig: rot quaoar}.

For our work, we also retrieved data obtained with the Hubble Space Telescope (HST) on four different dates in 2024 and 2025. These data are available in FITS format, with the data products accessible via the MAST archive \citep{Proudfoot_2025}. We downloaded the data corresponding to the following dates: 2024-05-21, 2025-03-26, 2025-03-29, and 2025-04-05. The images used had pixel units of electrons ($e^-$), and the flux in these units was divided by the exposure time. We did aperture photometry using PRAIA \citep{Assafin_2023} and to avoid contamination from nearby stars, we applied an aperture photometry correction \citep{Sirianni_2005}, adopting an aperture radius of 6 pixels for our measurements.

To calculate the magnitude, we utilized Equation \ref{eq: m_r}, where in this specific case $F$ is given by:
\begin{equation}\label{F mag}
    F = \frac{\text{counts}}{\text{EXPTIME} \times \text{CORR}{ab}}
\end{equation}
where $\text{counts}$ represents the total integrated signal measured within the photometric aperture, $\text{EXPTIME}$ is the total exposure time of the image in seconds, and $\text{CORR}{ab}$ is the aperture correction factor applied to account for the object's flux that falls outside our chosen 6-pixel radius.

Thus, using Equation \ref{eq: m_r}, we obtain the ST magnitude (STmag) through the following expression:
\begin{equation}
    \text{STmag} = m_r + \text{ZP}
\end{equation}
where $\text{STmag}$ is the magnitude in the Space Telescope (ST) system, which is based on a constant flux density per unit wavelength where the zero-point ($\text{ZP}$) is defined as:
\begin{equation}
    \text{ZP} = -2.5\log_{10}(\text{PHOTFLAM}) + \text{PHOTZPT}
\end{equation}
In this equation, $\text{PHOTFLAM}$ is the inverse sensitivity of the detector representing the flux density in that produces one count per second, and $\text{PHOTZPT}$ is the standard ST magnitude zero-point. Both keyword values are intrinsically calibrated and provided by the HST data products.

Finally, to obtain the magnitude used in this work, we converted the values to the AB magnitude system (ABmag):\begin{equation}
\text{ABmag} = \text{STmag} - 5\log_{10}(\text{PHOTPLAM}) + 18.6921
\end{equation}
where $\text{PHOTPLAM}$ is the pivot wavelength of the specific filter used during the observation. \citep{Sirianni_2005}.

Therefore, these two light curves are be employed in determining the triaxial model of Quaoar. To constrain the shape, we also utilized data from 37 stellar occultations by Quaoar observed between 2011 and 2025, which are available in \cite{Margoti_2026}.

Furthermore, since Quaoar possesses a ring system, it is highly unlikely that this system lies outside the object's equatorial plane \citep{Marzari_2020}. We utilized this physical constraint as a prior for the shape determination. If we do not constrain the pole position, the parameter space for Quaoar makes the axis ratios indeterminate. In particular, the most accurate and physically plausible configuration is one in which the ring system is purely equatorial \citep{Morgado_2021, Pereira_2023}. Therefore, Quaoar and its ring share the same pole.

The prior values we adopted for the pole position come from a new determination using unpublished ring detections (Pereira et al. 2026, in preparation). In this configuration, the pole is given by $\alpha_0 = 259.90 \pm 1.22^{\circ}$ and $\delta_0 = 55.12 \pm 0.36^{\circ}$ evaluated at a $5\sigma$ confidence level, which allows for a ring eccentricity of up to 0.01. These values are consistent with those reported by \citet{Proudfoot2025ring}.

The complete set of priors used for our modeling is summarized in Table \ref{tab: priors Quaoar}:

\begin{table}[htb!]
\caption {Priors of Quaoar model} \label{tab: priors Quaoar}
\centering
\begin{tabular}{lll}
\hline\hline             
Variable & Central value & range \\
\hline
Primary axis ($a$)       & 631 km & any $>$ $1$\\
Secondary axis ($b$)    & 550 km & any $<=$ $a$ \\
Polar axis ($c$)        & 510 km & any $<=$  $b$\\
rotational period ($P$) & 17.6728 h & 0.0062 h \\
Initial phase ($W_0$)   & 30$^{\circ}$ & 90$^{\circ}$  \\
Pole orientation ($\alpha_0$) & 259.90$^{\circ}$ & 1.22$^{\circ}$\\
Pole orientation ($\delta_0$) & 55.12$^{\circ}$  & 0.36$^{\circ}$\\
Center in a single event ($x_i,y_i$) & 0, 0 & 100, 100 \\
\hline
\end{tabular}
\end{table}

For the initial phase, we must establish a reference epoch. The choice of this date must be carefully selected. In the case of Quaoar, because we aim to assign greater statistical weight to the numerous, highly precise stellar occultations, we require a reference epoch that is temporally close to these events. Thus, we adopted 2022-08-09 06:34 UTC as our reference date. We prioritized the occultations not only for their high spatial precision but also because the available rotational light curves are temporally distant and lack comparable accuracy.

The stellar occultation data for Quaoar were extracted from \cite{Margoti_2026}, and we utilized the chords already projected onto the sky plane. The adopted coordinates correspond to Quaoar’s NIMA v19 ephemeris \citep{Desmars_2015} combined with the DE440 planetary ephemeris. There are a total of 214 points on Quaoar’s limb, accounting for all immersions and emersions. Consequently, in our modeling, we fit the global parameters $a$, $b$, $c$, $W_0$, $P$, $\alpha_0$, and $\delta_0$, alongside a local projected center ($f, g$) for each of the 37 occultation events, resulting in a total of 81 free parameters, 7 global and 74 local.

Regarding the model uncertainty ($\sigma_{model}$) introduced in Equation \ref{eq: chi2}, using the equivalent radius and density values derived by \cite{Margoti_2026} and assuming an undifferentiated icy object ($k=1$), the maximum topography Quaoar could theoretically support is $6.2 \pm 0.4$~km. Therefore, we could adopt a $\sigma_{model}$ of up to 6 km without violating physical constraints. However, we opted $\sigma_{model}$ of 2 km to remain consistent with the approach made in \cite{Margoti_2026}.

Furthermore, for each tested set of parameters, an optimal center offset ($f, g$) in the sky plane was determined. This search was restricted to a range of $\pm 100$ km, which corresponds to the maximum uncertainty of Quaoar's ephemeris plus an additional 10\% margin.

Finally, summarizing, for Quaoar's semi-axes, we allowed any positive values that respected the geometric continuity of a triaxial ellipsoid, i.e., $a \ge b \ge c$. The rotation period and its prior uncertainty were derived from the Lomb-Scargle analysis. The remaining parameters were allowed to explore their full valid parameter spaces, including the initial phase. 

Using the bounds for the semi-axes and initial phase outlined in Table \ref{tab: priors Quaoar}, we executed a run of the genetic algorithm described in Section \ref{sec: Modeling process} to identify the parameter set that best represents the Quaoar data. This optimal result, combined with the priors, was then utilized to initialize the MCMC walkers. The MCMC execution included a burn-in phase to thermalize the chains, followed by a main production phase that generated the posterior distributions.

\section{Results}\label{sec: Resultados}

In this section, we present the results of our modeling for Quaoar, as well as its implications for the dynamics of its ring and satellite system.

\subsection{Modeling}\label{sec: Modeling}

The simplest configuration that successfully reproduces Quaoar's rotational light curves and stellar occultations is a triaxial shape.  We used 295 photometric points from the \citet{Ortiz_2003} dataset, along with 32 points from the \citet{Proudfoot_2025} dataset, and combined them with 37 stellar occultations by Quaoar from \citet{Margoti_2026}, to constrain the model fit. This combined dataset comprises 108 chords (2 chords with 1 point), which corresponds to 214 individual points on Quaoar's limb. Although tested, we do not require a surface albedo variation to explain the amplitude of the observed rotational light curves (see appendix \ref{ap:albedo}).

For the \citet{Ortiz_2003} light curve, we assigned a photometric uncertainty ($\sigma$) of 0.018 mag for each point, while for the \citet{Proudfoot_2025} data, 0.002 mag was used. Each occultation chord possesses its own intrinsic observational uncertainty, and for the global model uncertainty ($\sigma_{model}$), we adopted 2 km as previously discussed.

The fitted variables are the equatorial semi-axes (primary $a$ and secondary $b$), the polar semi-axis ($c$), the pole position coordinates ($\alpha_0$ and $\delta_0$), the rotational period ($P$), and the initial phase ($W_0$). In addition to these global parameters, we also marginalized the astrometric center ($f_i, g_i$) for each individual occultation event. Each determined center offset generates a refined astrometric position for Quaoar; these corrected positions are presented in Appendix \ref{sec:New astrometric position}.

\begin{figure*}
    \includegraphics[scale=0.45]{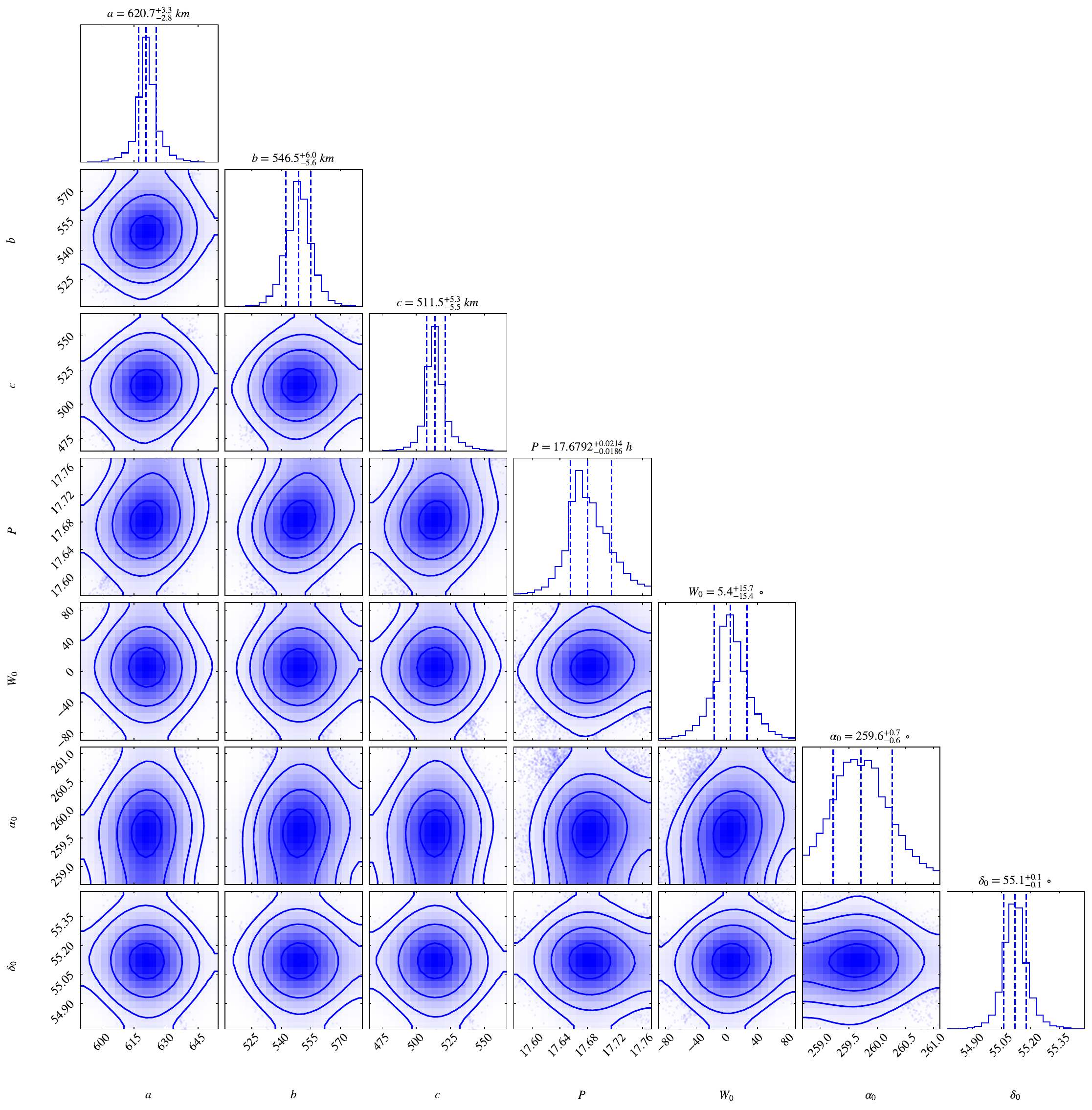}
    \caption{Corner plot showing the 1D marginalized and 2D joint posterior probability distributions for the parameters of the purely triaxial model. The diagonal panels display the 1D histograms for the semi-axes $a$ , $b$ , and $c$ , the rotation period $P$ , the initial phase $W_0$ , and the pole orientation coordinates $\alpha_0$ and $\delta_0$. The vertical dashed lines mark the median and the $1\sigma$ (16th and 84th percentiles) confidence intervals. The off-diagonal panels display the 2D contour plots illustrating the covariances between parameters.}
    \label{fig: corner plot}
\end{figure*}

The posterior distributions derived from our MCMC sampling are displayed in the corner plot in Figure \ref{fig: corner plot}. The minimum $\chi^2$ of the fit was 371.0, yielding a reduced $\chi^2$ (per degree of freedom) of 0.805. This result, below the one, indicates possible mild over-fitting to the data. The final results obtained for Quaoar's shape are summarized in Table \ref{tab:results}.

\begin{table}[h!]
    \caption{Summary results.}
    \centering
    \begin{tabular}{lc}
        \hline \hline
        Variable & Result\\
        \hline
        Equatorial semi-axis (a)       & $620.7^{+3.3}_{-2.8}$~km           \\
        Equatorial semi-axis (b)       & $546.5^{+6.0}_{-5.6}$~km           \\
        Polar semi-axis (c)            & $511.5^{+5.3}_{-5.5}$~km           \\
        equatorial oblateness          & $0.120 \pm 0.006$   \\
        Polar oblateness - a           & $0.176 \pm 0.005$                  \\
        Polar oblateness - b           & $0.064 \pm 0.003$                  \\
        $J_2$                          & $0.052 \pm 0.004$                   \\
        $J_4$                          & $-0.0065 \pm 0.0010$               \\
        $C_{22}$                       & $0.014 \pm 0.001$                  \\
        D$_{\text{e}}$ volumetric      & 1115.7~$\pm$~9.4~km               \\
        D$_{\text{e}}$ area            & 1118.6~$\pm$~9.3~km               \\
        Density                        & 1.663~$\pm$~0.127~g/cm$^3$         \\
        Rotational period ($p$)        &$17.6792^{+0.0214}_{-0.0186}$\\
        Initial phase ($W_0$)          &$5.4^{+15.7}_{-15.4}$\\
        Pole orientation ($\alpha_Q$)  & $259.6^{+0.7}_{-0.6}$     \\
        Pole orientation ($\delta_Q$)  & $53.1   \pm 0.1 {^{\circ}}$        \\
        Geometric albedo\footnote{The geometric albedo was obtained following the procedures presented by \citet{Margoti_2026}.}               & 0.111~$\pm$~0.012                   \\
        \hline
    \end{tabular}
    \label{tab:results}
\end{table}

\begin{figure*}[htb!]
\gridline{\fig{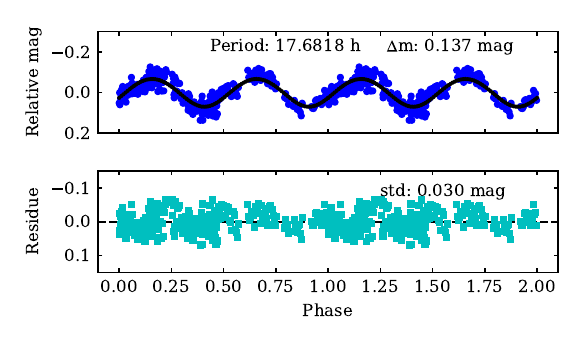}{0.5\textwidth}{(a)}
          \fig{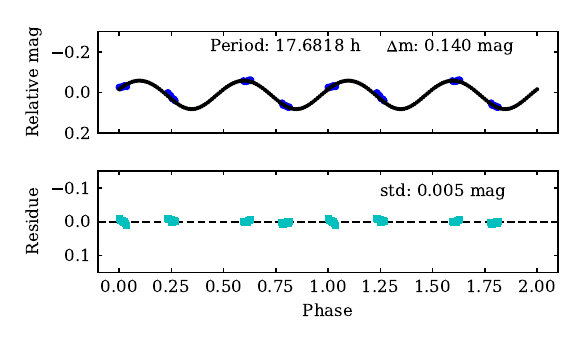}{0.5\textwidth}{(b)}
          }
\caption{Rotational light curves of Quaoar folded to the best-fit double-peak period of $P = 17.6818$ h. Panel (a) shows the photometric data acquired in 2003, while Panel (b) presents the HST observations. In both panels, the top sub-plot displays the observed relative magnitudes (blue circles) alongside the synthetic light curve generated by the triaxial shape model (black solid line). The data points include their respective observational uncertainties, as discussed in the text. The bottom sub-plots show the residuals of the fit (cyan squares) and their corresponding standard deviations. Both light curves are plotted over two full rotational phases for better visualization of the double-peaked features.}
\label{fig: rot quaoar model}
\end{figure*}

The best-fit model superimposed on the rotational data is shown in Figure \ref{fig: rot quaoar model}. In this figure, we highlight that the variation in Quaoar's aspect angle between 2003 and 2024 caused a nominal difference of only 0.003 in the model's amplitude ($\Delta mag$). Furthermore, the standard deviation of the residuals (data minus model) demonstrates consistency, yielding values comparable to the optimal fit obtained via the Lomb-Scargle periodogram.

\begin{figure*}[htb!]
    \centering
    \includegraphics{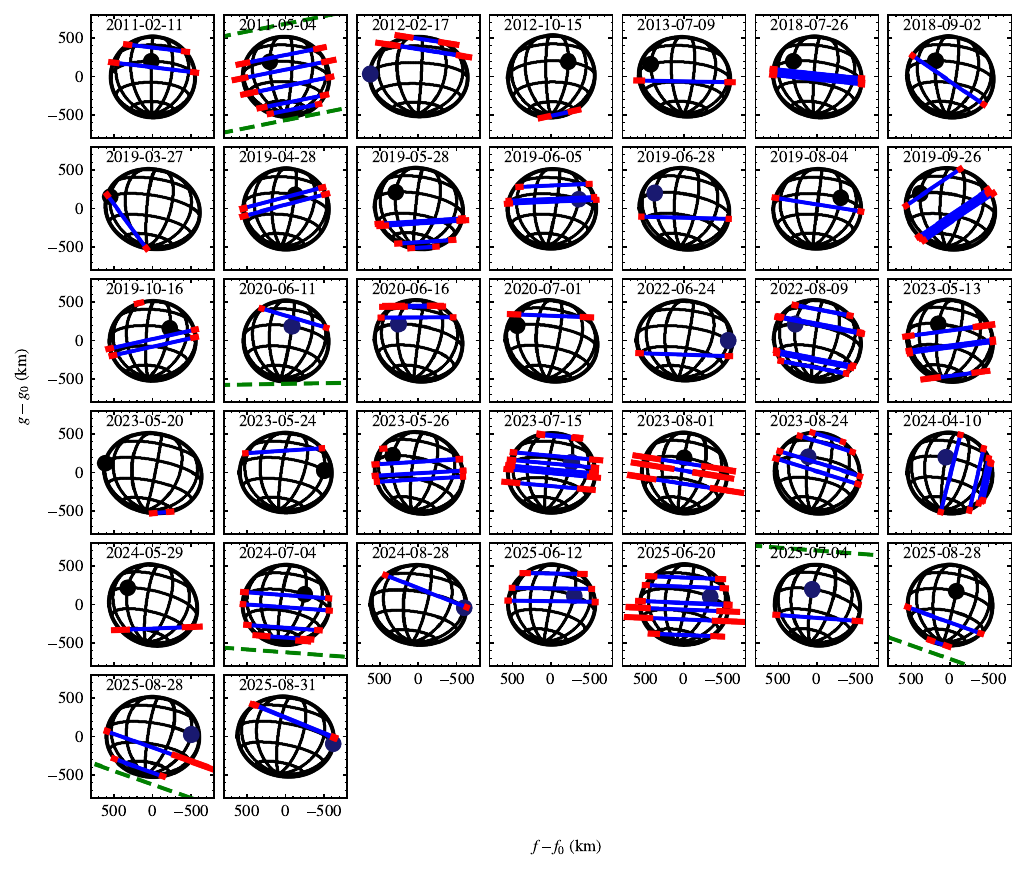}
    \caption{Best-fit triaxial shape model of Quaoar projected onto the sky plane across the multiple stellar occultation events observed between 2011 and 2025. Each panel corresponds to a specific epoch, displaying the apparent limb of the model against the observations. The solid blue lines represent the positive occultation chords, while the red lines indicate the uncertainties at their respective extremities. The dashed lines denote the closest negative chords, which provided upper limits to constrain the physical extent of the model. The axes, $f - f_0$ and $g - g_0$, represent the local sky-plane coordinates in kilometers relative to the fitted center for each event showed in Appendix \ref{sec:New astrometric position}.}
    \label{fig: occ quaoar}
\end{figure*}

\begin{figure*}[htb!]
    \centering
    \includegraphics[scale=0.7]{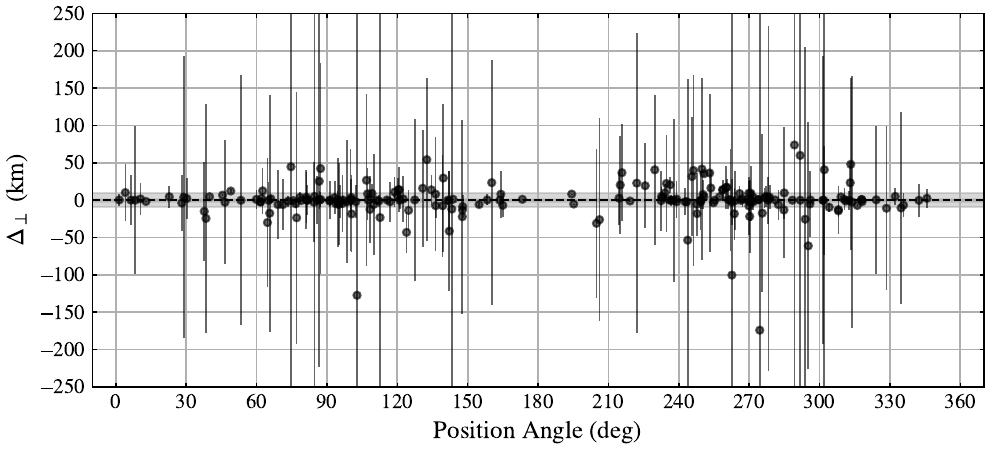}
    \caption{Perpendicular limb residuals ($\Delta_{\perp}$) of Quaoar as a function of the position angle. The data points represent the differences between the observed occultation chord extremities and the projected 3D shape model. The residuals were projected along the normal direction to the local tangent of the modeled limb. The horizontal dashed line denotes the zero-residual reference, indicating a perfect match with the theoretical limb, while the shaded grey ($\sim 9$ km) area represents the model's uncertainty, convolved with the adopted tolerance and the median uncertainty of the fitted center.}
    \label{fig: resid occ}
\end{figure*}

The respective limb fits for the occultation chords are displayed in Figure \ref{fig: occ quaoar}. Furthermore, Figure \ref{fig: resid occ} presents the normal distance of each occultation chord with respect to the mean-limb. Without considering the model uncertainty, the error bars of 34 chord extremities (16\% of the total) do not cross the horizontal dotted line at zero, which represents the theoretical limb of the modeled Quaoar. However, by incorporating our model tolerance ($\sigma_{model}$) alongside the shape model uncertainties listed in Table \ref{tab:results}, which corresponds to approximately 9 km, we delineate the shaded region shown in Figure \ref{fig: resid occ}. When accounting for this region, the number of non-intersecting extremities drops to just 5 (or 2\%).

In Figure \ref{fig: resid occ}, the standard deviation of the points relative to this zero baseline is 25.8 km, while the weighted standard deviation is 2.1 km. These low residual values demonstrate the high quality of the shape determination, especially given the large number of integrated data points.

\subsection{Weywot's Mass and Bulk Density} \label{weywotmass}

The astrometry derived from stellar occultations provides the coordinates of the target's center at the epoch of each event. However, standard numerical integrations and orbital ephemeris for binary Trans-Neptunian systems, such as the NIMA framework \citep{Desmars_2015}, track the center of mass (barycenter) of the system rather than the primary body alone. Because Weywot has a non-negligible mass, Quaoar orbits the system's shared barycenter, resulting in a small but measurable astrometric wobble projected onto the sky plane \citep{Liberato_2026}.

The displacement of Quaoar's center of figure relative to the system's barycenter is a direct consequence of the gravitational pull exerted by its satellite. By mapping the cross-track ($f$) and along-track ($g$) astrometric offsets between the occultation, derived geometric center and the barycentric ephemeris, we can isolate this reflex motion. Quaoar's barycentric displacement is linearly proportional to Weywot's relative orbital position ($dx, dy$) scaled by the mass ratio of the system. Therefore, the observed offsets can be modeled as:
\begin{equation}\label{eq: lon, lat, r}
    \begin{split}
        f = q \cdot \delta x + f_o; \\
        g = q \cdot \delta y + g_o, \\
    \end{split}
\end{equation}
where $q = M_{\text{Weywot}} / M_{\text{Quaoar}}$ is the mass ratio, and $f_o$ and $g_o$ account for residual systemic offsets between the reference star catalog (e.g., Gaia DR3) and the ephemeris. This linear dependence allows us to dynamically determine the system's mass ratio, and, consequently, Weywot's mass.

The primary observables for our barycentric motion analysis are the astrometric offsets of Quaoar's geometric center relative to the reference ephemeris. These cross-track and along-track offsets, denoted as $f$ and $g$ respectively, were determined from the spatial fitting of the stellar occultation chords to the 3D shape model (see Apendix \ref{sec:New astrometric position}). To correlate Quaoar's astrometric wobble with the gravitational pull of its satellite, we require Weywot's relative position on the sky plane ($\delta x, \delta y$) at the epoch of each occultation. For the majority of our dataset, these relative coordinates and their inherent orbital uncertainties were retrieved using the Miriade ephemeris web service, with the orbital fit of \citet{Braga_2025}.

During two specific occultation campaigns, on 2019-08-04 and 2023-05-26, the observing geometry and chord distributions allowed for the simultaneous positive detection of both Quaoar and Weywot. For these two specific epochs, rather than relying on the theoretical ephemeris, we adopted the relative separation ($\delta x, \delta y$) directly from the observational data. 

To determine the mass ratio $q$, we performed a simultaneous Ordinary Least Squares (OLS) regression on both the cross-track ($f$) and along-track ($g$) components. To comprehensively account for the observational error, the inherent positional uncertainties associated with the Weywot's orbit were projected onto the astrometric offsets (the y-axis of our model). Upon analyzing the initial astrometric residuals, we opted to preserve all of our dataset to capture the natural astrometric scatter, excluding only a single extreme geometric outlier from the 2018-09-02 event (which exhibited anomalous offsets of $f \approx -522$ km and $g \approx -462$ km). Removing this specific data point was sufficient to yield a good linear fit.

Because Quaoar's barycentric displacement is anti-correlated with Weywot's relative position, the physical mass ratio $q$ corresponds directly to the absolute value of the fitted negative slope. Once the slope is established, Weywot's mass is straightforwardly calculated as $M_{\text{Weywot}} = q \times M_{\text{Quaoar}}$. For this calculation, we adopted Quaoar's mass of $(1.208 \pm 0.063) \times 10^{21}$ kg \citep{Braga_2025}.

Finally, to provide constraints on Weywot's physical composition, we derived its bulk density by dividing the calculated mass by its assumed spherical volume, adopting an effective diameter of 116-172 km \citep{Valenzuela_2026}. The final uncertainties for both the satellite's mass and its bulk density were determined through standard analytical error propagation, summing in quadrature the statistical standard error of the OLS slope, the uncertainty in Quaoar's primary mass, and the considerable uncertainty in Weywot's radius.

\begin{figure*}
    \centering
    \includegraphics[scale=0.75]{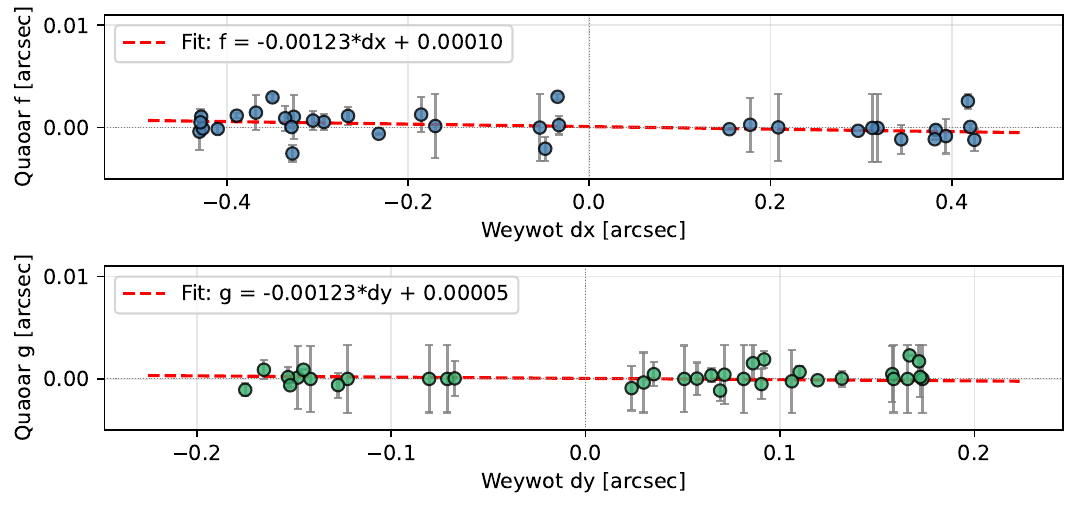}
    \label{fig: weywot mass}
    \caption{Quaoar's barycentric displacement induced by Weywot. The panels display the cross-track ($f$ vs. $\delta x$, left) and along-track ($g$ vs. $\delta y$, right) astrometric offsets in arcseconds. Green circles represent the center-of-ellipsoid positions derived from stellar occultations. The gray error bars represent the total uncertainty in the y-axis, which combines the uncertainty of Quaoar's fitted center of ellipsoid with the uncertainty of Weywot's relative separation projected onto the y-axis. The red dashed line denotes the simultaneous OLS linear fit applied to the retained dataset. The absolute value of the resulting slope ($\vert{}-0.00123\vert{}$) directly yields the system's mass ratio $q$.}
\end{figure*}

The linear regression of the barycentric astrometric offsets yielded a best-fit slope of -0.0012 $\pm$ 0.0003, see figure \ref{fig: weywot mass}. Taking the absolute value of this slope, we derived a mass ratio of $q$ = 0.0012 $\pm$ 0.0003 for the Quaoar-Weywot system. Applying this ratio to the primary's established mass, we dynamically determined Weywot's mass to be (1.485 $\pm$ 0.360) $\times 10^{18}$ kg. Furthermore, by combining this mass estimate with the adopted effective diameter, we computed a bulk density of 0.95 $\pm$ 0.60 g/cm$^3$ for the satellite. 

\textcolor{black}{It is worth noting that traditional unresolved astrometry (e.g., Gaia DR3 epoch astrometry) is susceptible to photocenter shifts induced by surface albedo variegations, shape elongation, or the unresolved flux of a satellite. The expected photocenter wobble induced by Weywot is on the order of a few milliarcseconds, whereas Gaia DR3 single-epoch uncertainties for this $V \sim 19$ target range from $\sim90$ to $\sim600$ mas, completely burying the signal in noise. In contrast, stellar occultations directly probe the physical limb of the target with precisions often better than 2 mas. Consequently, the geometric centers derived from our 3D shape model represent the true center of figure.}

\subsection{Dynamical implications to Quaoar's environment}\label{sec:environment}
\begin{figure*}[htb!]
\centering
\gridline{
        \fig{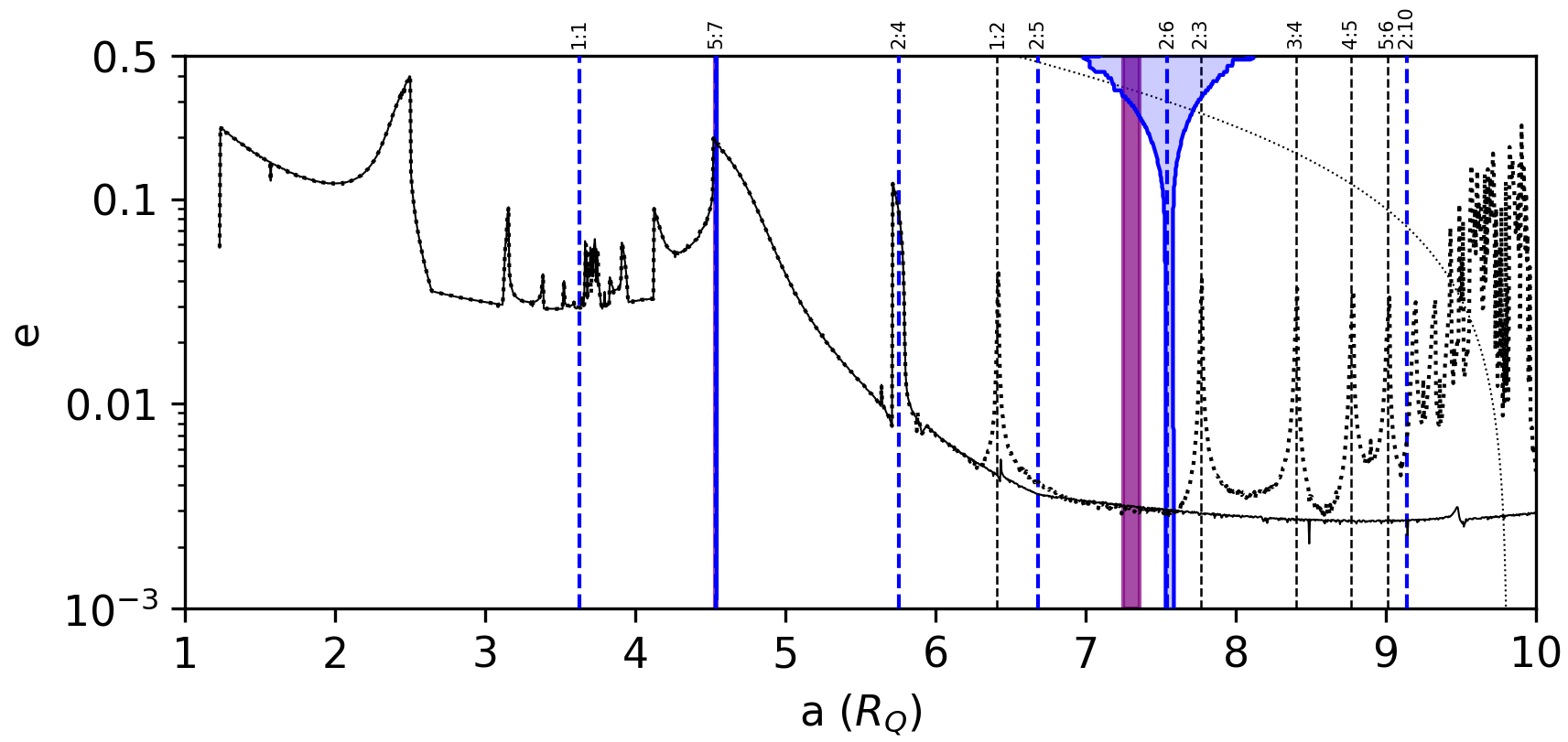}{0.5\textwidth}{(a)\label{dynamics1}}
        \fig{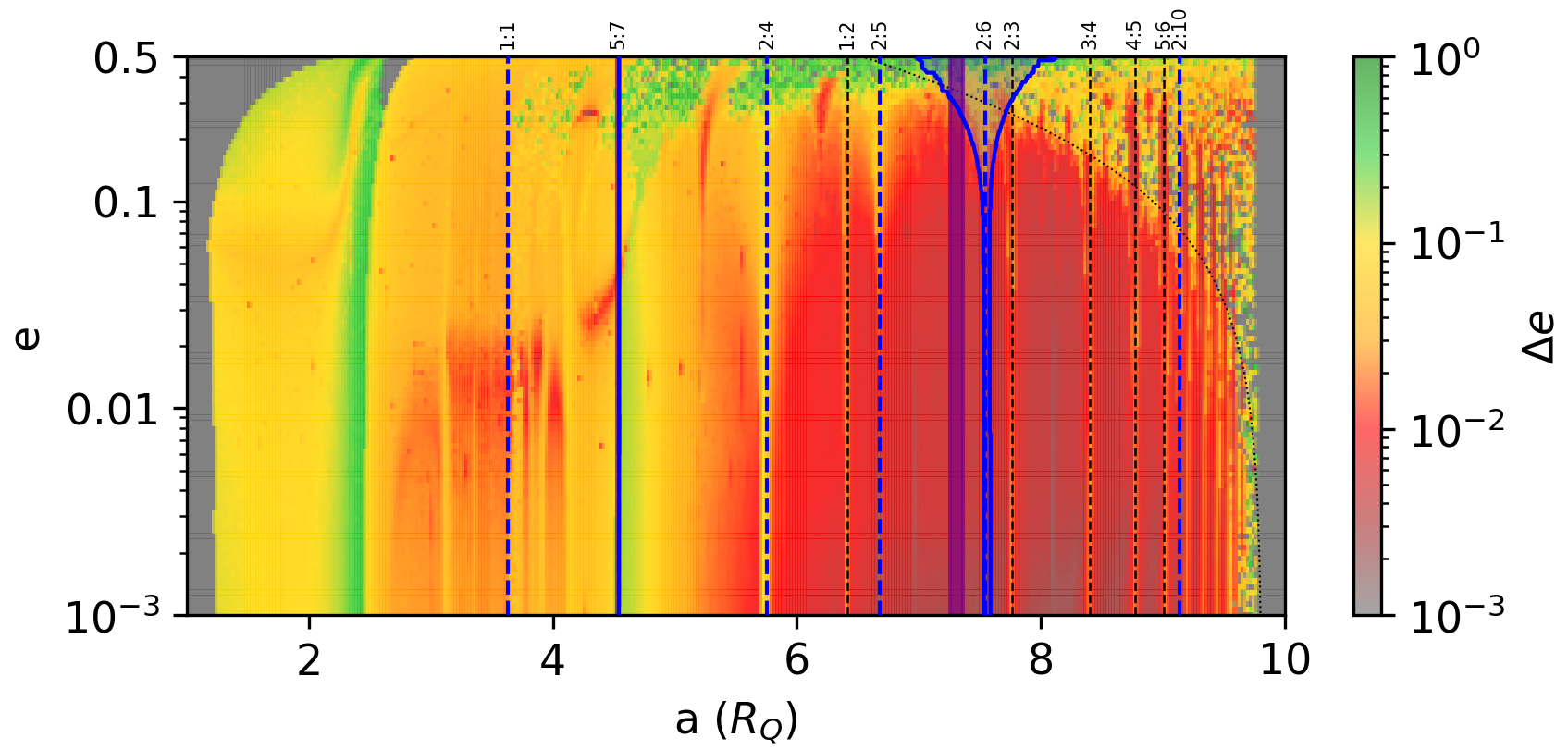}{0.5\textwidth}{(b)\label{dynamics2}}
    }
\caption{Dynamical environment around Quaoar obtained using the triaxial shape derived in this work. Panel (a) shows the mean eccentricity induced in initially nearly circular particles by Quaoar’s triaxial gravitational field and Weywot, with (solid line) and without (dotted line) the recently detected feature, assuming it to be a satellite. Panel (b) shows the eccentricity variation, $\Delta$e, as a function of the particles’ initial semimajor axis and eccentricity for the complete dynamical model, including triaxial Quaoar, Weywot, and the feature. In both panels, vertical blue dashed lines indicate the locations of the relevant spin-orbit resonances, while black dashed lines indicate mean-motion resonances with the feature. The blue shaded regions mark the regions in which particles remain trapped in the 2:6 and 5:7 spin-orbit resonances.}
    \label{fig:dynamics}
\end{figure*}

The shape derived in this work has direct consequences for the dynamics of material orbiting Quaoar. As systematically explored in previous works, the rotating non-axisymmetric gravitational field of a triaxial object strongly affects the surrounding material. First, the equatorial asymmetry of the central body perturbs orbiting trajectories, inducing non-zero osculating eccentricities \citep{Winter2019,Ribeiro2021,Ribeiro2023,Beauge2025,Madeira2025,Madeira2026}. Where the orbital motion becomes commensurable with the rotation of the central body, spin-orbit resonances (SORs) emerge and may promote radial confinement \citep{Sicardy2020,Sicardy2026,Salo2026,Celletti2026}. Finally, sufficiently close to the primary, the overlap of low-order SORs and the strong non-axisymmetric perturbation can drive chaotic eccentricity diffusion, leading to collisions with Quaoar or escape from the system \citep{GiuliattiWinter2023,Madeira2025,Madeira2026}.

To assess these effects using the new shape model of Quaoar, we performed a set of numerical experiments using the N-body code \texttt{REBOUND} \citep[for details, see][]{Madeira2025}, including Quaoar, Weywot, and a set of massless test particles. The orbital parameters of Weywot were taken from \citet{Braga_2025}, while its physical parameters were adopted from the results presented in Section~\ref{weywotmass}. For Quaoar, we adopted the values of $J_2$ and $C_{22}$ listed in Table~\ref{tab:results}, computed under the assumption of a homogeneous interior, together with a rotation period of $P=17.6792$~h and a volumetric-equivalent radius of $R_Q=557.8$~km. Some simulations also included the putative satellite recently reported in the system, adopting the parameters of \citet{Ribas_2026}. Each particle was integrated for 10,000 orbital periods. 

Our results are summarized in Figure~\ref{fig:dynamics}. Panel (a) shows the mean eccentricity acquired by initially circular and planar particles, with and without the putative satellite. Panel (b) shows the total eccentricity variation, $\Delta e$, as a function of the particles' initial osculating semimajor axis and eccentricity for the full system including Quaoar, Weywot, and the putative satellite. In both panels, the particles are initialized at pericenter, with longitude of pericenter $\varpi=90^{\circ}$, in order to sample the equilibrium configuration associated with the 2:6 SOR \citep{Roig2026}. The locations and widths of the 2:6 and 5:7 SORs, proposed to be associated with \textcolor{black}{Quaoar's main ring} (Q1R) and \textcolor{black}{secondary ring} (Q2R), respectively, were obtained by analysing the libration of the corresponding resonant angles \citep[see][]{Madeira2026}.

The innermost region shows strong eccentricity excitation ($e \gtrsim 0.1$), with material interior to $\sim 2.4~R_Q$ completely depleted in less than 3 years. This limit therefore provides a dynamical inner boundary for the possible population of small satellites proposed by \citet{Ribas_2026}. In addition, the mean eccentricity acquired by initially circular particles provides a dynamical eccentricity floor for possible small satellites in this region.

The eccentricity excitation obtained with the new shape model is stronger than that found in previous dynamical studies of the Quaoar system \citep[e.g.][]{Madeira2026}. This is consistent with the stronger equatorial asymmetry of the new shape, which yields $C_{22}=0.014$, approximately three times the previous estimate of $C_{22}=0.005$ \citep{Margoti_2024}. The region interior to $\sim8~R_Q$ is dynamically dominated by Quaoar's triaxial shape, with particles between $\sim2.4~R_Q$ and Q2R reaching mean eccentricities of $\sim0.03$-$0.2$. Beyond $\sim6~R_Q$, non-resonant particles typically maintain eccentricities of $\sim0.003$. The putative satellite introduces localized eccentricity enhancements near its MMRs, which constitute its main dynamical effect in this region. As expected, the satellite also efficiently clears its surrounding chaotic zone (gray region in Figure~\ref{dynamics2}). Consistent with \citet{Madeira2026}, Weywot has only a minor effect on particle eccentricities in the inner Quaoar system ($a \lesssim 10~R_Q$).

\begin{figure*}[htb!]
\centering
\gridline{
        \fig{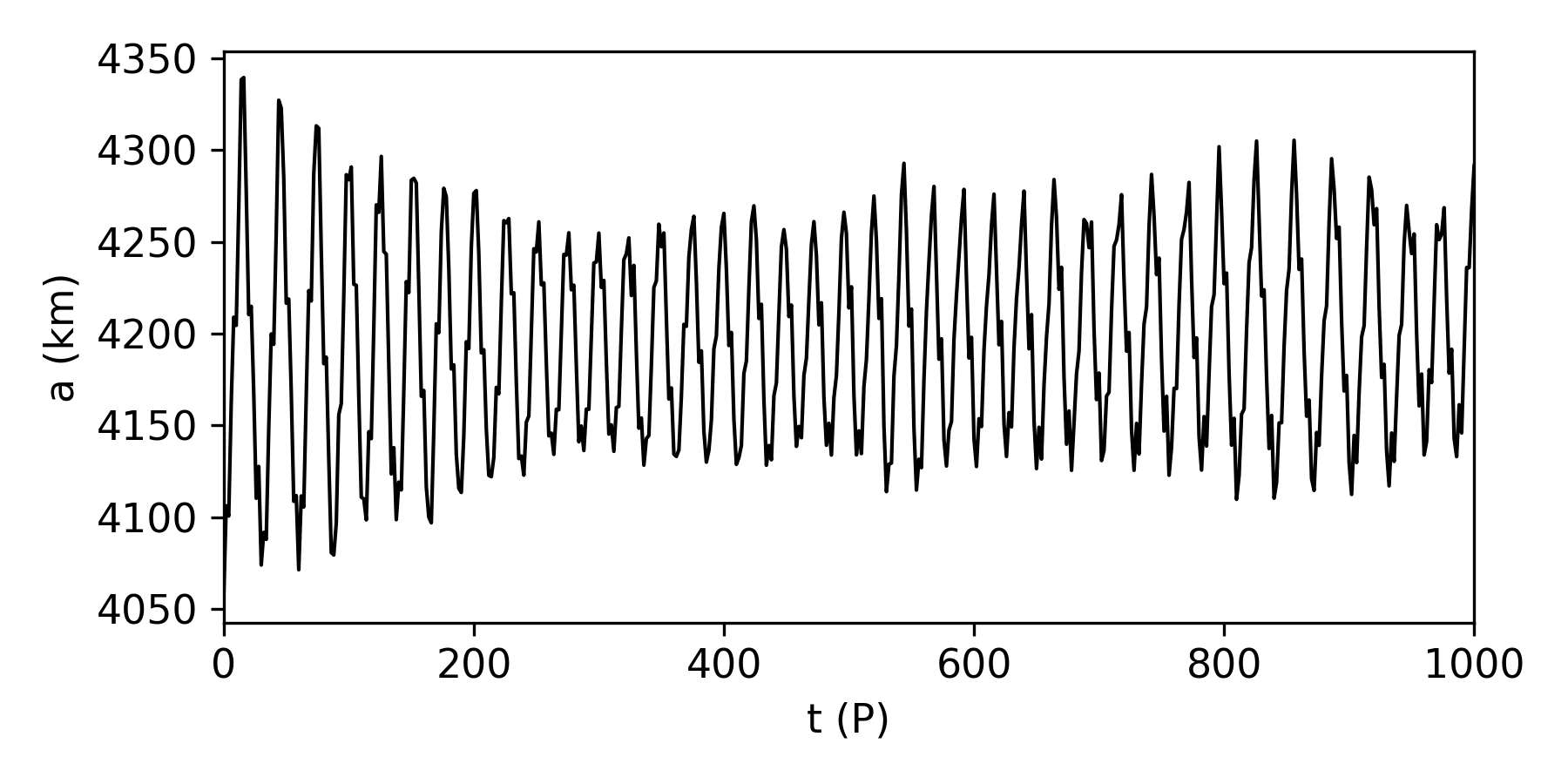}{0.5\textwidth}{(a)\label{dynamics3}}
        \fig{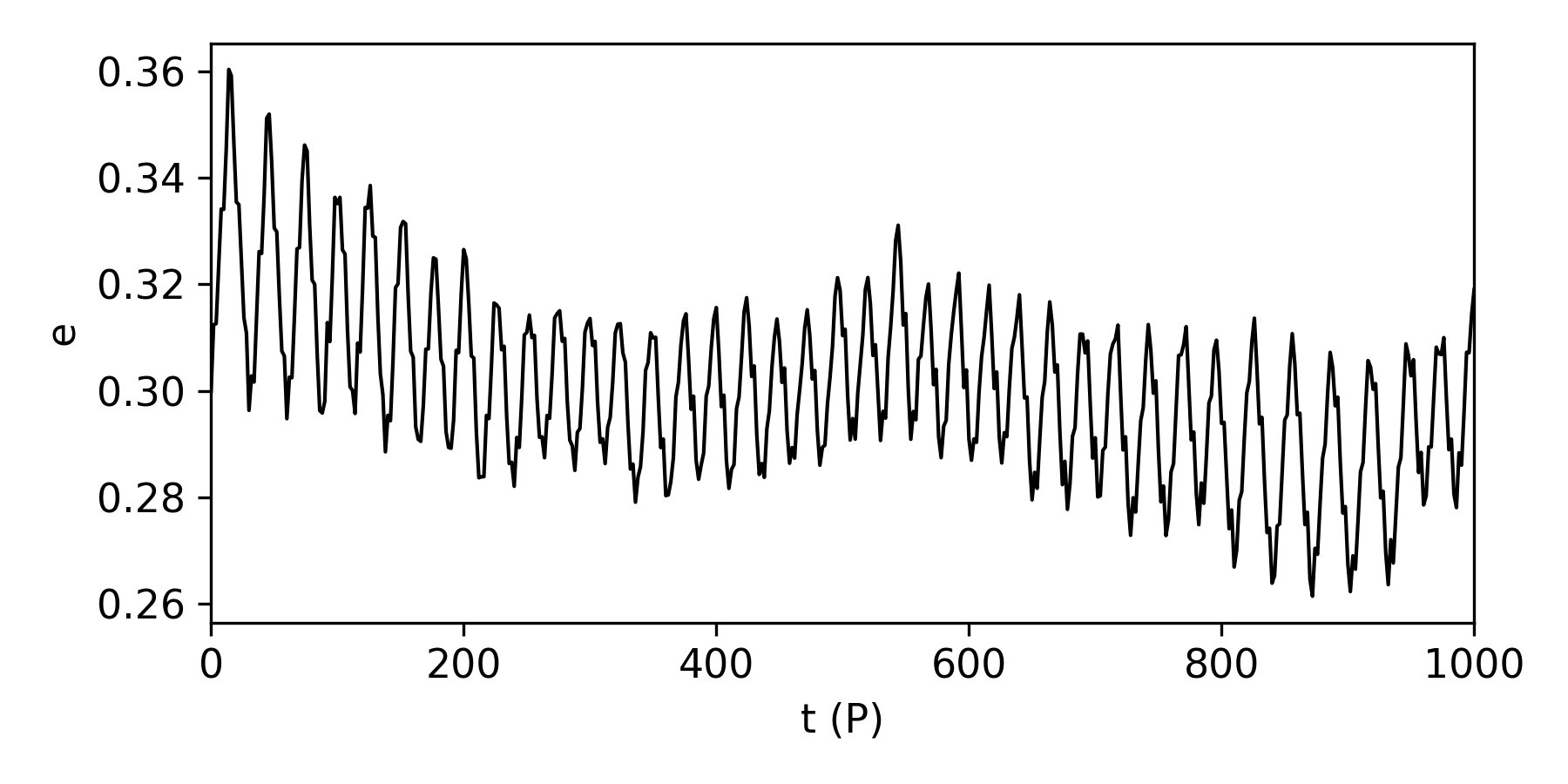}{0.5\textwidth}{(b)\label{dynamics4}}
    }
\gridline{
        \fig{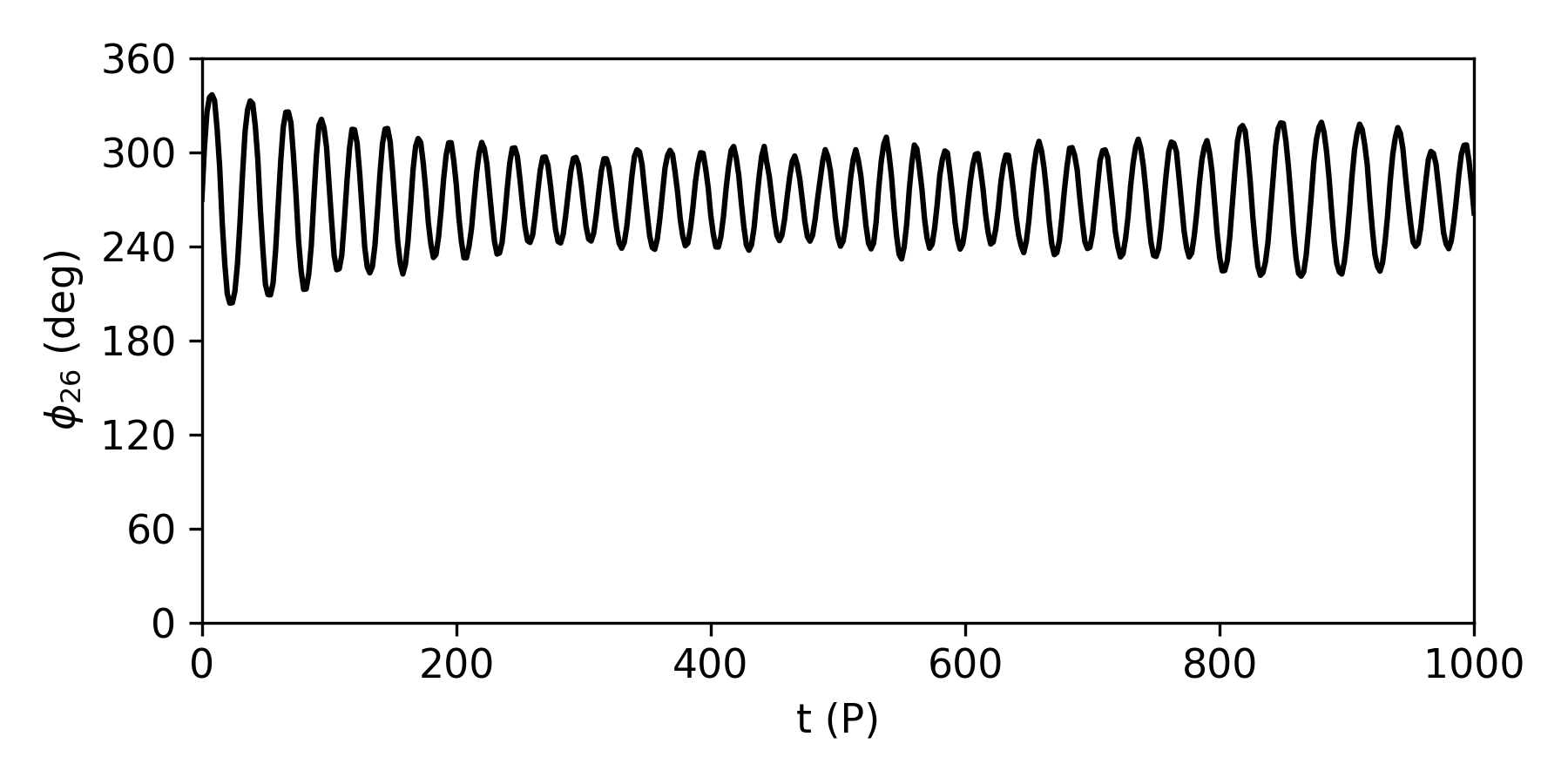}{0.5\textwidth}{(c)\label{dynamics5}}
        \fig{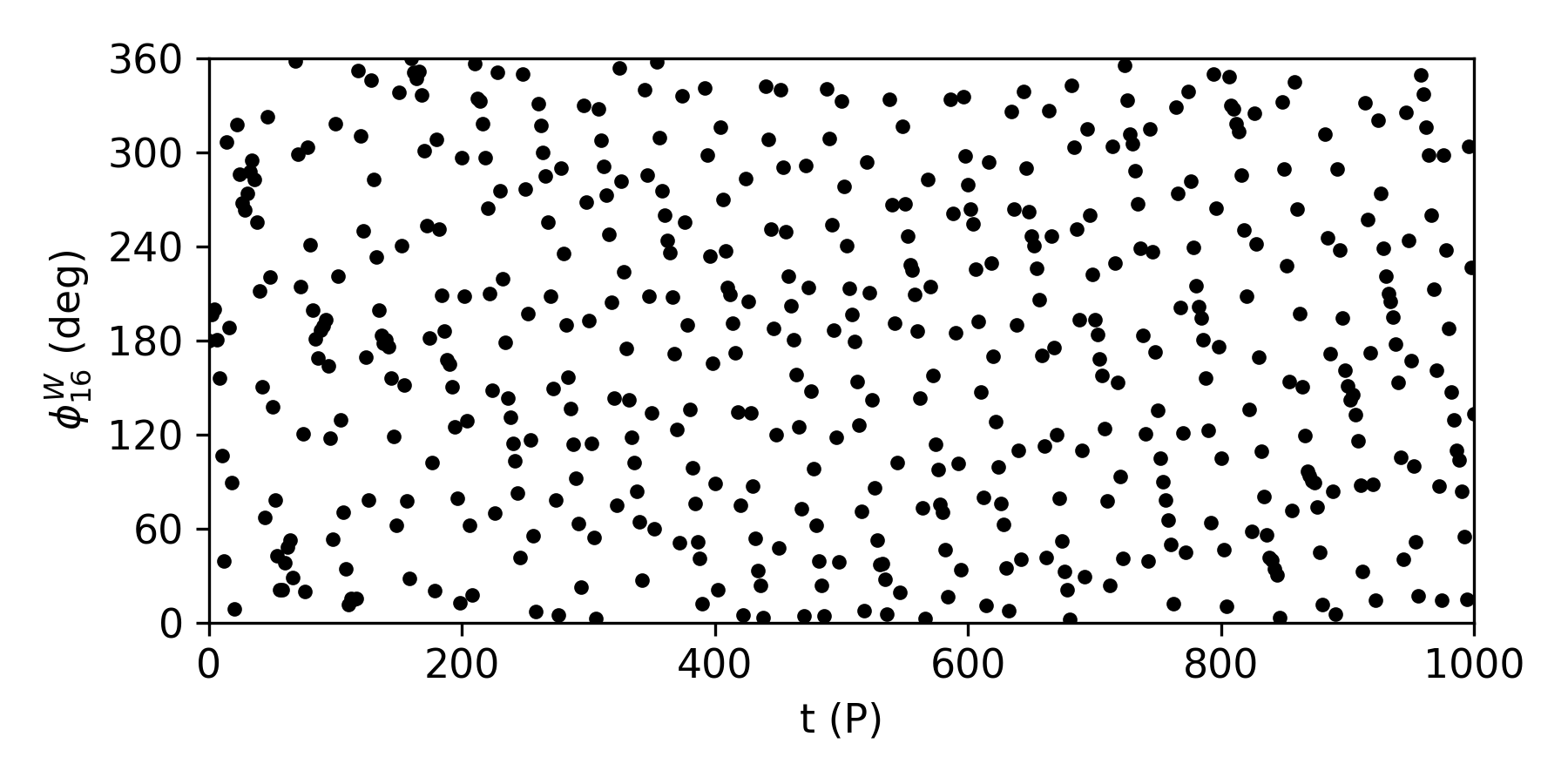}{0.5\textwidth}{(d)\label{dynamics6}}
    }

\caption{Temporal evolution of the orbital elements and resonant angles of a particle initially placed at the radial centre of Q1R with an initial eccentricity of $e_0=0.3$. Panels (a) and (b) show the evolution of the semimajor axis and eccentricity, respectively. Panels (c) and (d) show the resonant angles associated with the 2:6 spin-orbit resonance with Quaoar and the 1:6 mean-motion resonance with Weywot, respectively.}
    \label{fig:phiQ1R}
\end{figure*}

Focusing on the rings, we find that both Q1R and Q2R lie within dynamically stable regions. The 5:7 SOR closely matches the location of Q2R, with a separation of less than $20$~km between the resonance and ring centers, assuming Q2R to be located at $2529$~km \citep{Proudfoot2025ring}. Simulations of particles initialized within Q2R show libration of the corresponding resonant angle, demonstrating that particles in the ring region can be trapped in the 5:7 SOR and supporting a possible role of this resonance in the confinement of Q2R. These particles exhibit an eccentricity floor of $\sim0.05$ in the collisionless simulations, which can be taken as a reference value for the ring eccentricity. In a real ring, however, collisional damping is expected to modify this value.

The 2:6 SOR center extracted from our numerical simulations, located at $\sim4202$~km, shows a larger offset from the Q1R center at $4096$~km. Nevertheless, at eccentricities $\gtrsim0.2$, the resonance width overlaps the radial location of Q1R, indicating that the ring could still be associated with the resonance if its particles occupy sufficiently eccentric orbits. Interestingly, such eccentric orbits may also provide a possible explanation for the strong azimuthal variation observed in the radial width of Q1R \citep{Morgado_2023,Pereira_2023}. If Q1R is approximated by apsidally aligned elliptical streamlines, its radial width naturally varies with longitude \citep[see ][]{French1986}. The relatively high eccentricities required for overlap with the 2:6 SOR could therefore help explain the observed contrast between the minimum and maximum ring widths. 

Quaoar’s size and density suggest that it likely underwent internal melting and at least partial differentiation \citep{Emery2024}. The resulting central concentration of denser rocky material would yield values of $J_2$ and $C_{22}$ lower than those estimated here for a homogeneous interior \citep{Tricarico2014}. For example, \citet{Proudfoot_2025} showed that complete differentiation could reduce Quaoar’s $J_2$ by up to 40\% relative to the homogeneous case, with the magnitude of this reduction depending on the core radius. This reduction would shift the location of the 2:6 SOR \citep{Ribeiro2023}, thereby modifying its radial overlap with Q1R. However, we do not expect this shift to qualitatively alter our conclusions, since the resonance location obtained with the homogeneous model is already close to the nominal location derived solely from the commensurability between the orbital and rotational periods, corresponding to the spherical limit.

The evolution of a Q1R particle with an initial eccentricity of $e_0=0.3$ is shown in Figure~\ref{fig:phiQ1R}. The libration of the 2:6 SOR angle, $\phi_{26}$, demonstrates that, at high eccentricities, particles within the Q1R region can indeed be trapped in this resonance. In contrast, the 1:6 MMR angle with Weywot, $\phi^{W}_{16}$, associated with a resonance previously proposed to affect the ring \citep{Morgado_2023}, circulates at the nominal Q1R location. For the same initial eccentricity, however, $\phi^{W}_{16}$ librates for particles with semimajor axes of $\sim4062$~km. This suggests that the 1:6 MMR may still influence part of the Q1R region, potentially producing azimuthal clumps that could contribute to the large asymmetries observed in the ring optical depth \citep{Rodriguez2023}. Dedicated simulations are required to test this hypothesis. Overall, the results of this section show that the revised triaxial shape changes the dynamical picture of Quaoar's inner system, strengthening the possible association of both rings with SORs, while the effects of Weywot and the putative satellite remain predominantly local.

\section{Conclusions}\label{sec:conclusions}

In this work, we developed a methodology capable of deriving an ellipsoidal shape model for any object, utilizing either a single type of data, as previously demonstrated by \citet{Margoti_2026}, or a combination of different observational techniques, as presented in this paper. Reconstructing the 3D shape of TNOs is notoriously difficult due to their slow orbital motion and minimal aspect angle variation over human timescales. 

By simultaneously modeling 37 stellar occultation events obtained between 2011 and 2025 alongside two independent photometric datasets (from 2003 and recent HST observations in 2024 and 2025), we applied this framework to characterize the large Trans-Neptunian Object (50000) Quaoar. Furthermore, by mapping each section of our modeled ellipsoid with planetocentric latitudes and longitudes, we were able to parameterize localized spot regions. This allows us to generate and test models with albedo variations across any geometric configuration.

Our MCMC analysis demonstrates that Quaoar is best represented by a triaxial shape with equatorial semi-axes $a = 620.7^{+3.3}_{-2.8}$ km and $b = 546.5^{+6.0}_{-5.6}$ km, and a polar semi-axis $c = 511.5^{+5.3}_{-5.5}$ km. This shape naturally produces a double-peaked rotational light curve with a period of $P = 17.6792^{+0.0214}_{-0.0186}$ h, matching both the 2003 and the HST photometric amplitudes. 

While an oblate shape harboring a massive albedo spot can mathematically reproduce the light curve (as explored in Appendix \ref{ap:albedo}), the required physical extent of such a spot renders it highly improbable. Therefore, the purely triaxial model stands as the simplest and most physically sound solution.

This 3D shape determination yields a volumetric equivalent radius of $557.8 \pm 4.7$ km (diameter of $1115.7 \pm 9.4$ km) and a bulk density of $1.663 \pm 0.127$ g cm$^{-3}$, which, being larger in volume than the model by \citet{Margoti_2026}, results in a lower density. It also provides an albedo of 0.111~$\pm$~0.012.

By isolating Quaoar's barycentric motion from the astrometric offsets derived from the stellar occultations, we determined the mass of its satellite, Weywot. The simultaneous linear regression of the cross-track and along-track residuals yielded a system mass ratio of $q = 0.0012 \pm 0.0003$. Applying this ratio to the primary's established mass, we obtained a satellite mass of $(1.485 \pm 0.360) \times 10^{18}$ kg. Combined with Weywot's adopted effective volume, this results in a bulk density of $0.95 \pm 0.60$ g cm$^{-3}$.

From a dynamical perspective, the revised figure of Quaoar, together with the Weywot parameters derived in this work, has direct consequences for the architecture of the inner system. Compared with previous shape models, the stronger equatorial asymmetry produces larger eccentricity excitation, with Quaoar's non-axisymmetric gravitational field dominating the dynamics within $\sim8~R_Q$. Material interior to $\sim2.4~R_Q$ is rapidly removed, providing a dynamical inner limit for any putative population of small prograde satellites, while the effects of Weywot and the recently reported feature, assuming it to be a satellite, remain predominantly local. Both rings lie within dynamically stable regions and show a strong connection with spin-orbit resonances: Q2R closely coincides with the 5:7 SOR, whereas Q1R overlaps the 2:6 SOR for eccentricities $\gtrsim0.2$. These relatively high eccentricities may  contribute to the pronounced azimuthal variation observed in the radial width of Q1R.

As a natural byproduct of the shape modeling, we marginalized the center of each stellar occultation to minimize the radial residuals. This process yielded a set of new astrometric positions for Quaoar at the epochs of the events, which are all tied to the same global shape model. These positions, corrected for the gravitational light deflection caused by the Sun, Jupiter, and Saturn, are provided in Appendix \ref{sec:New astrometric position}. With these positions derived from the triaxial model the determination of the rings' physical parameters, assuming circular rings that share Quaoar's center can improve. Furthermore, it enables an event-by-event comparison with the oblate model presented by \citet{Margoti_2026}. These updated astrometric constraints also will be pivotal for future orbital redeterminations of Quaoar, ensuring the continuous predictability of stellar occultations by its main body, its ring, and its satellite.

This work presents a consistent methodology for determining shape models of distant objects that exhibit minimal variation in aspect angle and are large enough that their topographic variations are not highly relevant. Although it is possible to apply this framework to inner Solar System objects, it is not recommended due to the simplicity of the model, as these objects have much more comprehensive observational datasets \citep{Gradovski_2026}.

The application to Quaoar represents a step forward in our knowledge of this object. Until now, the best model for Quaoar was an oblate model \citep{Margoti_2026, Ribas_2013}. The strongest point of agreement between Quaoar's oblate model and the data is precisely the density associated with an object in Maclaurin hydrostatic equilibrium. This yields a bulk density very similar to the density derived from Weywot's orbital mass and the theoretical density of a Maclaurin object, as presented by \citet{Margoti_2026}, in addition to better agreement regarding the stable regions for the rings \citep{Madeira2026}.

Although the oblate model fits the occultation chords well, it relies on an albedo spot model with serious issues to match the rotational light curves. The first issue is the sheer size of this spot, which would cover nearly half of a hemisphere; such a feature is difficult to produce naturally, as the mechanisms required for its formation would have to be extreme. Furthermore, to represent both the 2003 and HST rotation curves within this oblate framework, we are forced to consider different rotation periods. While these periods still fall within the error bars, this discrepancy represents a statistical issue.

For our triaxial model, these problems are not present. The triaxial shape naturally reproduces the rotational light curves, and the rotation period remains the same for both datasets. Additionally, the normal distances between the occultation points and the triaxial model are smaller than those presented by \citet{Margoti_2026}, yielding an improvement in matching the apparent oblateness of the projected ellipses across the individual occultations. Furthermore, by assuming a differentiated body, we can still reproduce a stable region where ring particles do not require high eccentricities. This places them near the 1/3 SOR, which appears to be the most dynamically relevant resonance when discussing ring systems around small bodies \citep[e.g.,][]{Sicardy_2024, Ortiz_2017, Morgado_2023, Pereira_2023, Ribas_2026}.

\section{Acknowledgments}
This study was financed in part by the Coordenação de Aperfeiçoamento de Pessoal de Nível Superior - Brasil (CAPES) - Finance Code 001.
This work was partly funded by the Spanish projects PID2020- 112789GB-I00 (AEI) and Proyecto de Excelencia de la Junta  de Andalucía PY20-01309. Financial support from the Severo Ochoa grant CEX2021-001131- S (MCIN/AEI/10.13039/501100011033) is also acknowledged. 
BS ackowledges support by the French ANR project Roche, number
ANR-23-CE49-0012.

\section{Data availability}

The re-reduced 2003 photometric data, the recent HST light curves, and the occultation chords projected onto the sky plane (based on the NIMAv19 + DE440 ephemerides) can be publicly accessed at \dataset[https://doi.org/10.5281/zenodo.21185812]{https://doi.org/10.5281/zenodo.23417651}\footnote{or in the \textit{GitHub} repository \url{https://github.com/giuliano-margoti/Quaoar---data.git}}.

\textcolor{black}{All the HST data used in this paper were obtained from the Mikulski Archive for Space Telescopes (MAST) and can be formally accessed via \citet{https://doi.org/10.17909/3xvd-zh90}.}

\appendix

\section{\textbf{Oblate model with albedo variation}}\label{ap:albedo}

As demonstrated by \citet{Margoti_2026}, Quaoar can be well-represented by an oblate shape that successfully fits the occultation chords. In this appendix, we utilize the framework developed in the methodology section to generate a synthetic light curve for Quaoar, considering an albedo variation on an oblate body. 

Detailed knowledge regarding albedo or shape variations among outer Solar System objects remains scarce, with only a handful of bodies having these features well-determined. The Pluto-Charon system is particularly rich in this regard. Pluto possesses a large and bright region covering approximately $1.6 \times 10^6$ km$^2$, known as Tombaugh Regio \citep{Stern_2015}. Charon is similar in size to Quaoar and also features a dark spot at its pole, named Mordor Macula, whose formation is likely linked to gases escaping from Pluto \citep{Grundy_2016_2}. This spot covers roughly $3 \times 10^5$ km$^2$, corresponding to about 6\% of Charon's total surface area.

Therefore, albedo variations might be common among large objects like Quaoar. Rotational light curves indicative of albedo variations have also been observed for Haumea, which is believed to harbor a reddish spot covering roughly 20\% of its surface \citep{Lacerda_2008}. Curiously, Haumea also possesses a ring system \citep{Ortiz_2017}, along with two satellites \citep{Brown_2005}, similar to the Quaoar system. (84522) 2003~VS$_2$ also exhibits a probable albedo variation, as its light curve displays a double peak with distinctly different amplitudes \citep{Rossi_2019, Lubiano_2022}.

We investigated the characteristics of a surface spot on Quaoar that could reproduce the rotational light curve presented in Figure \ref{fig: rot quaoar}c. To fit the spotted model, we applied the same procedures described in Section \ref{sec: Modeling process}, introducing four new variables to parameterize the albedo variation: the planetocentric latitude ($\lambda$) and longitude ($\beta$) of a circular spot, its angular radius ($r$) in degrees, and a brightness factor (\textit{fat}), representing the brightness of a spot compared to the rest of the body.

The prior values utilized for the MCMC fit follow the same logic as before and are summarized in Table \ref{tab: albedo_priors}. For these priors, we adopted the semi-axes $a$, $b$, and $c$, along with the pole orientation, as given by \citet{Margoti_2026}. We used the single-peak rotation period derived in this work (see Table \ref{tab: rotational period by Quaoar}) and set the initial phase fix to zero. This allows the spot's location to vary freely across the entire body, with the latitude ranging from $-90^\circ$ to $90^\circ$ and the longitude from $0^\circ$ to $360^\circ$. To avoid geometric degeneracy, the spot's radius was strictly limited to a maximum of $90^\circ$ (a hemisphere). A radius larger than $90^\circ$ is mathematically equivalent to placing an inverted-brightness spot at the antipodal point. The brightness factor (\textit{fat}) was constrained between 1 (same brightness as the body) and 10. To model a darker spot without losing generality, one could simply place it at the antipodal position with a fractional brightness factor ($0 < fat < 1$).

\begin{table}[htb!]
\caption{Priors of the spotted Quaoar model} 
\label{tab: albedo_priors}
\centering
\begin{tabular}{lll}
\hline
\hline
Variable & Central value & Range \\ 
\hline
Primary axis ($a$)       & 566.1 km & +2.5, -2.2\\
Secondary axis ($b$)     & $=a$ & Fixed \\
Polar axis ($c$)         & 511.2 km & +3.6, -3.7\\
Rotational period ($P$) & 8.8364 h & $\pm$ 0.0031 h \\
Initial phase ($W_0$)   & 0$^{\circ}$ & Fixed  \\
Pole orientation ($\alpha_0$) & 259.7$^{\circ}$ & $\pm$ 0.2$^{\circ}$\\
Pole orientation ($\delta_0$) & 53.4$^{\circ}$  & $\pm$ 0.2$^{\circ}$\\
Albedo Latitude ($\lambda$) & 0$^{\circ}$ & $\pm$ 90$^{\circ}$\\
Albedo Longitude ($\beta$) & 180$^{\circ}$ & $\pm$ 180$^{\circ}$\\
Albedo radius ($r$) & 45$^{\circ}$ & 0$^{\circ}$ -- 90$^{\circ}$\\
Albedo factor (\textit{fat}) & 6 & $\pm 4$ \\
\hline
\end{tabular}
\end{table}

\begin{figure*}[htb!]
    \centering
    \includegraphics{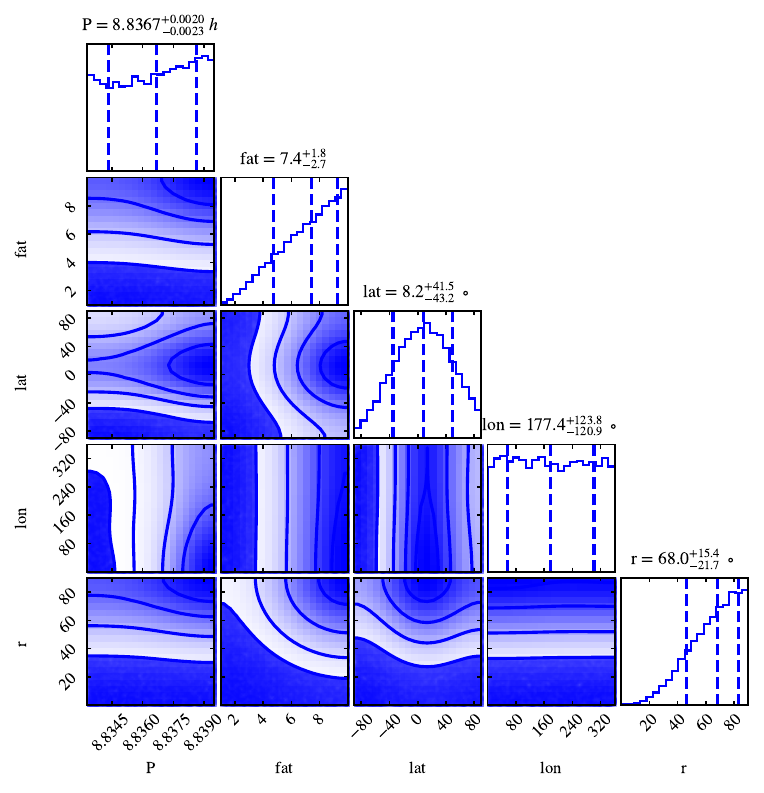}
    \caption{Corner plot showing the 1D marginalized and 2D joint posterior probability distributions for the parameters of the spotted oblate model. The diagonal panels display the 1D histograms for the single-peak rotation period ($P$), the spot brightness factor ($fat$), its planetocentric latitude ($lat$) and longitude ($lon$), and its angular radius ($r$). The vertical dashed lines mark the median and the $1\sigma$ (16th and 84th percentiles) confidence intervals. The off-diagonal panels display the 2D contour plots illustrating the covariances between the albedo parameters.}
    \label{fig: corner plot albedo}
\end{figure*}

The MCMC results indicate that the spot can be located at essentially any longitude, as the imposed limits on $a$, $b$, $c$, $P$, and the pole orientation allow the parameter to span the entire search space. However, as expected, there is a strong preference for a location near the equator, facing the observer's line of sight. The brightness factor remains somewhat unconstrained within the searched range, though it shows a preference for values between 5 and 10. Because the spot radius is truncated at $90^\circ$, the MCMC posterior distributions show that coherent solutions require a spot radius of at least $40^\circ$. A smaller radius would produce a prominent flat plateau in the synthetic light curve, which contradicts the continuous variation seen in the \citet{Ortiz_2003} data. The posteriors for the spot parameters (period, latitude, longitude, radius, and brightness factor) are shown in the corner plot in Figure \ref{fig: corner plot albedo}.

\textcolor{black}{It is important to emphasize that the MCMC exploration for this spotted oblate model remains poorly constrained. As shown in the posterior distributions (Figure \ref{fig: corner plot albedo}), the spot's angular radius does not converge to a bounded value within the explored interval. Instead, the highest probability saturates at the $90^\circ$ upper limit, which corresponds to a full hemisphere. This lack of formal convergence highlights the intrinsic degeneracy of fitting a localized albedo feature using solely light curves from a nearly constant aspect angle. Without additional observational constraints, the exact size, location, and contrast of the spot remain largely undetermined, which further reinforces the pure triaxial model as the more mathematically robust and physically sound solution.}

\begin{figure*}[htb!]
    \centering
    \includegraphics{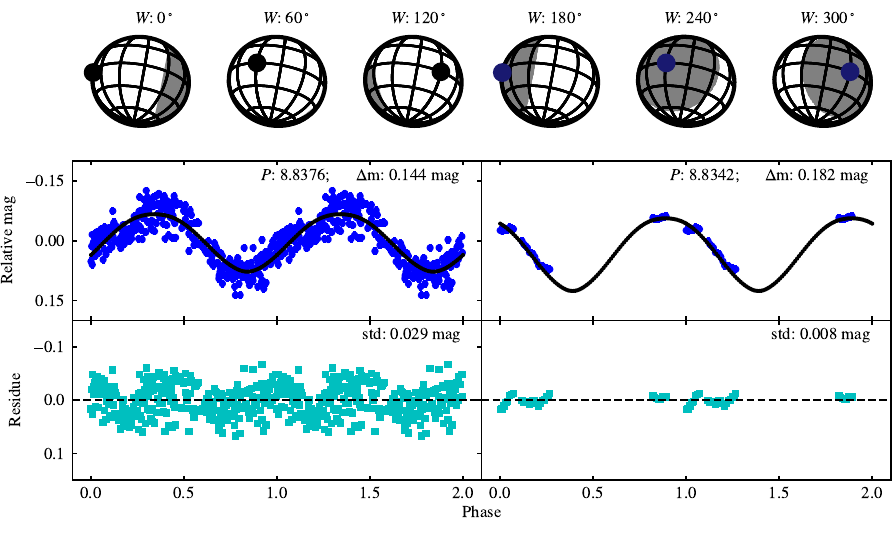}
    \caption{Synthetic light curves and physical representation of the spotted oblate model for Quaoar. The top sequence illustrates the 3D oblate model and the position of its surface albedo spot at various rotational phases ($W$). The middle panels display the 2003 photometric data (left) and the HST observations (right) as blue circles, superimposed with the synthetic light curves (black solid lines) generated by the spot-driven single-peak model. For this plot slightly different rotation periods were adopted: $P = 8.8376$ h for the 2003 data and $P = 8.8342$ h for the HST data. The bottom panels show the residuals of the fits (cyan squares) alongside their standard deviations.}
    \label{fig: rot quaoar albedo}
\end{figure*}

When comparing the nominal results to the rotational data, the model fits the 2003 observations exceptionally well. However, it exhibits a poorer fit to the HST data. While this might initially suggest that the nominal oblate model is unfavorable, introducing minor adjustments to the rotation period, well within its formal uncertainties, yields satisfactory results for the HST light curve (Figure \ref{fig: rot quaoar albedo}), indicating that the oblate shape with an albedo spot remains a viable geometric solution.

Despite providing a viable fit, a spot covering almost half of Quaoar's surface is physically somewhat unsatisfying in a broader context. It is highly improbable for such a massive, coherent albedo feature to form, for comparison, the prominent spots on Pluto and Charon cover a much smaller fraction of their respective surfaces. The formation mechanisms for such an albedo variation could be diverse. It might have resulted from a collision that excavated deeper materials, exposing a brighter icy interior. Although it is not known how widespread surface excavation by impacts is in the TNOs region, there is evidence of large craters on objects within this population. For example, Arrokoth, visited by the New Horizons spacecraft, features a prominent crater named Maryland, approximately 7 km wide on its smaller lobe \citep{Stern_2019}. Moreover, stellar occultation data for the TNO (307261) Mani 
revealed a depression over 40 km deep, interpreted as an impact crater \citep{Rommel_2023}. While the physical crater itself might not be large or deep enough to cause a discernible geometric variation in the light curve, given the photometric precision of the 2003 data, the excavated and ejected bright material could. This scenario is particularly compelling for Quaoar's environment, as a large impact would provide a natural mechanism for the formation of its ring system \citep{Morgado_2023}, its known satellite Weywot \citep{Braga_2025}, and potentially an extended debris or satellite belt \citep{Ribas_2026}. However, considering the sheer scale of the spot required by our MCMC models, an impact origin becomes physically improbable.

Another possibility is the past existence of a localized transient atmosphere on Quaoar that eventually collapsed and condensed back onto the surface, creating a region of higher albedo. This is similar to the mechanism proposed for Eris, where the collapse of a global atmosphere left its surface with an albedo very close to 1 \citep{Sicardy_2011}. Alternatively, ring material could have rained back down onto Quaoar's surface, accumulating in a preferential region and altering its local reflectivity.

\section{\textbf{New astrometric position}}\label{sec:New astrometric position}

In this appendix, we present the updated astrometric positions of Quaoar derived from the stellar occultations modeled in this study. For each event, the fitted center of the ellipsoidal model yields Quaoar's offset relative to the NIMAv19 ephemeris on the tangent plane, with the event geometry anchored to the Gaia DR3 coordinates of the occulted stars. We corrected these positions for the gravitational light deflection exerted by the Sun, Jupiter, and Saturn by adopting the values and procedures provided in Appendix C of \citet{Margoti_2026}, following the methodology of Poiani et al. (2026, in preparation). The final astrometric coordinates, along with the corrected $f$ and $g$ sky-plane offsets, are listed in Table \ref{tab:astrometric position}.

\begin{longtable}{ccccc}
\caption{This table contains Quaoar's astrometric positions at the time of closest approach for each stellar occultation. These positions have been corrected for the gravitational deflection caused by the Sun, Jupiter and Saturn.\label{tab:astrometric position}} \\
\hline
\multicolumn{1}{l}{\textbf{Event}} &
\multicolumn{1}{l}{\textbf{\shortstack[l]{{\color{black}f + df}\\ {\color{black}(km)}}}} &
\multicolumn{1}{l}{\textbf{\shortstack[l]{{\color{black}g + dg}\\ {\color{black}(km)}}}} &
\multicolumn{1}{l}{\textbf{\shortstack[l]{R.A. \\ h m s $\pm$ mas}}} &
\multicolumn{1}{l}{\textbf{\shortstack[l]{Decl. \\ d m s $\pm$ mas}}} \\
\hline
\endfirsthead
\multicolumn{3}{c}{\tablename\ \thetable{} -- Continuation} \\
\hline
\multicolumn{1}{l}{\textbf{Event}} &
\multicolumn{1}{l}{\textbf{\shortstack[l]{{\color{black}f + df}\\ {\color{black}(km)}}}} &
\multicolumn{1}{l}{\textbf{\shortstack[l]{{\color{black}g + dg}\\ {\color{black}(km)}}}} &
\multicolumn{1}{l}{\textbf{\shortstack[l]{R.A. \\ h m s $\pm$ mas}}} &
\multicolumn{1}{l}{\textbf{\shortstack[l]{Decl. \\ d m s $\pm$ mas}}} \\
\hline
\endhead
2011-02-11 10:05:49.940 & -12.4 $\pm$ 57.4 &  -28.6 $\pm$ 69.0 & 17 28 47.6192147 $\pm$ 1.819 & -15 41 58.929994 $\pm$ 2.190 \\
2011-05-04 02:38:30.680 & -7.0 $\pm$ 13.5 &  10.4 $\pm$ 20.7 & 17 28 50.7999825 $\pm$ 0.463 & -15 27 42.812867 $\pm$ 0.700 \\
2012-02-17 04:30:44.960 & 40.3 $\pm$ 55.1 &  14.5 $\pm$ 86.0 & 17 34 21.8438964 $\pm$ 1.748 & -15 42 10.328753 $\pm$ 2.727 \\
2012-10-15 00:41:18.820 & 5.0 $\pm$ 100.0 &  3.9 $\pm$ 96.2 & 17 28 10.1273219 $\pm$ 3.171 & -15 36 23.292760 $\pm$ 3.056 \\
2013-07-09 02:40:03.760 & 8.5 $\pm$ 81.0 &  -0.2 $\pm$ 100.0 & 17 34 40.4636231 $\pm$ 2.652 & -15 23 37.482333 $\pm$ 3.274 \\
2018-07-26 15:08:21.300 & 32.7 $\pm$ 15.8 &  -10.8 $\pm$ 89.4 & 18 00 39.4079201 $\pm$ 0.526 & -15 21 58.514578 $\pm$ 2.933 \\
2018-09-02 18:16:51.600 & -522.6 $\pm$ 79.7 &  -462.5 $\pm$ 88.7 & 17 59 02.0595090 $\pm$ 2.858 & -15 27 30.256897 $\pm$ 3.127 \\
2019-03-27 06:56:33.480 & -25.7 $\pm$ 52.7 &  58.9 $\pm$ 24.7 & 18 12 50.7835337 $\pm$ 1.836 & -15 27 52.104634 $\pm$ 1.014 \\
2019-04-28 06:47:02.040 & 32.5 $\pm$ 65.7 &  -0.2 $\pm$ 100.0 & 18 12 23.2136998 $\pm$ 2.346 & -15 22 07.269227 $\pm$ 3.398 \\
2019-05-28 23:40:49.480 & -34.8 $\pm$ 42.2 &  1.0 $\pm$ 23.6 & 18 10 38.9338328 $\pm$ 1.406 & -15 18 37.534669 $\pm$ 0.831 \\
2019-06-05 03:00:57.040 & -18.2 $\pm$ 2.7 &  27.1 $\pm$ 28.6 & 18 10 07.3114853 $\pm$ 0.367 & -15 18 12.211936 $\pm$ 1.037 \\
2019-06-28 12:35:40.720 & -2.4 $\pm$ 8.8 &  0.0 $\pm$ 99.7 & 18 08 15.5284497 $\pm$ 0.311 & -15 17 57.989001 $\pm$ 3.288 \\
2019-08-04 17:24:51.320 & 15.9 $\pm$ 39.1 &  -0.2 $\pm$ 100.0 & 18 05 34.7868793 $\pm$ 1.288 & -15 21 02.068373 $\pm$ 3.279 \\
2019-09-26 18:55:33.920 & -4.8 $\pm$ 1.7 &  27.8 $\pm$ 2.0 & 18 04 27.3892044 $\pm$ 0.139 & -15 30 01.779593 $\pm$ 0.165 \\
2019-10-16 01:39:00.780 & 17.3 $\pm$ 25.8 &  21.1 $\pm$ 17.7 & 18 05 07.9984476 $\pm$ 0.837 & -15 33 24.860600 $\pm$ 0.593 \\
2020-06-11 16:32:13.560 & 21.0 $\pm$ 28.3 &  5.2 $\pm$ 30.7 & 18 15 03.0558556 $\pm$ 0.935 & -15 15 04.934882 $\pm$ 1.015 \\
2020-06-16 08:38:20.160 & -36.2 $\pm$ 33.5 &  0.8 $\pm$ 47.7 & 18 14 40.8009463 $\pm$ 1.125 & -15 15 02.167178 $\pm$ 1.595 \\
2020-07-01 21:38:55.520 & 0.2 $\pm$ 100.0 &  -0.2 $\pm$ 100.0 & 18 13 25.4861260 $\pm$ 3.320 & -15 15 23.732644 $\pm$ 3.331 \\
2022-06-24 12:09:29.520 & 27.8 $\pm$ 37.2 &  -0.2 $\pm$ 100.0 & 18 25 04.0903266 $\pm$ 1.312 & -15 07 59.527208 $\pm$ 3.346 \\
2022-08-09 06:34:02.600 & -34.6 $\pm$ 1.2 &  -4.1 $\pm$ 0.9 & 18 21 42.8677545 $\pm$ 0.243 & -15 12 45.829850 $\pm$ 0.221 \\
2023-05-13 08:40:33.160 & 34.8 $\pm$ 26.3 &  -7.3 $\pm$ 92.9 & 18 33 31.7144383 $\pm$ 0.905 & -15 04 52.024161 $\pm$ 3.059 \\
2023-05-20 00:24:21.460 & -1.0 $\pm$ 100.0 &  1.4 $\pm$ 50.9 & 18 33 09.7160077 $\pm$ 3.311 & -15 04 16.055048 $\pm$ 1.728 \\
2023-05-24 14:36:05.100 & 7.3 $\pm$ 27.6 &  5.7 $\pm$ 60.8 & 18 32 52.8645938 $\pm$ 0.942 & -15 03 56.444831 $\pm$ 2.026 \\
2023-05-26 15:53:07.440 & 35.5 $\pm$ 6.7 &  13.9 $\pm$ 35.2 & 18 32 44.9285508 $\pm$ 0.295 & -15 03 49.081620 $\pm$ 1.175 \\
2023-07-15 00:16:17.840 & 89.5 $\pm$ 10.9 &  -34.6 $\pm$ 30.5 & 18 28 56.6541235 $\pm$ 0.906 & -15 05 12.672945 $\pm$ 1.464 \\
2023-08-01 22:18:44.900 & 0.8 $\pm$ 100.0 &  -0.1 $\pm$ 100.0 & 18 27 39.0728406 $\pm$ 3.724 & -15 07 33.063777 $\pm$ 3.662 \\
2023-08-24 03:22:10.460 & 1.7 $\pm$ 3.9 &  -19.1 $\pm$ 6.0 & 18 26 27.7618136 $\pm$ 1.130 & -15 11 20.245614 $\pm$ 0.405 \\
2024-04-10 03:58:50.740 & -9.5 $\pm$ 1.7 &  70.6 $\pm$ 0.4 & 18 39 51.8965563 $\pm$ 0.152 & -15 04 33.771674 $\pm$ 0.179 \\
2024-05-29 19:15:55.320 & -0.8 $\pm$ 100.0 &  -0.2 $\pm$ 100.0 & 18 37 58.1146687 $\pm$ 4.057 & -14 58 37.852398 $\pm$ 3.655 \\
2024-07-04 22:57:30.880 & 2.1 $\pm$ 8.8 &  -15.4 $\pm$ 43.3 & 18 35 10.6992790 $\pm$ 0.752 & -14 59 34.749941 $\pm$ 1.607 \\
2024-08-28 10:38:30.040 & 44.8 $\pm$ 51.7 &  -0.2 $\pm$ 99.8 & 18 31 41.2280777 $\pm$ 2.059 & -15 07 52.299088 $\pm$ 3.488 \\
2025-06-12 09:53:32.740 & -3.9 $\pm$ 13.0 &  -18.3 $\pm$ 36.8 & 18 42 29.3680645 $\pm$ 0.591 & -14 53 02.400148 $\pm$ 1.341 \\
2025-06-20 23:46:41.500 & 91.1 $\pm$ 8.6 &  51.7 $\pm$ 48.1 & 18 41 49.2831136 $\pm$ 1.529 & -14 53 18.759558 $\pm$ 2.068 \\
2025-07-04 23:06:15.360 & -76.2 $\pm$ 23.7 &  12.4 $\pm$ 87.4 & 18 40 41.6337280 $\pm$ 1.083 & -14 54 18.399751 $\pm$ 3.011 \\
2025-08-28 02:44:10.100 & -62.8 $\pm$ 35.7 &  -32.6 $\pm$ 19.7 & 18 37 10.4092755 $\pm$ 1.539 & -15 02 54.335147 $\pm$ 1.446 \\
2025-08-31 13:46:22.600 & 79.0 $\pm$ 22.7 &  46.7 $\pm$ 53.3 & 18 37 03.1506799 $\pm$ 1.024 & -15 03 35.759890 $\pm$ 1.895 \\

\hline
\end{longtable}

\bibliography{sample631}{}
\bibliographystyle{aasjournal}



\end{document}